%% file: 19508.tex
\documentclass{aa}  
\usepackage{graphicx}
\usepackage{txfonts}
\newcommand{\mincir}{\raise -2.truept\hbox{\rlap{\hbox{$\sim$}}\raise5.truept
\hbox{$<$}\ }}
\newcommand{\magcir}{\raise -2.truept\hbox{\rlap{\hbox{$\sim$}}\raise5.truept
\hbox{$>$}\ }}
\newcommand{\siml}{\raise -2.truept\hbox{\rlap{\hbox{$\sim$}}\raise5.truept
\hbox{$<$}\ }}
\newcommand{\simg}{\raise -2.truept\hbox{\rlap{\hbox{$\sim$}}\raise5.truept
\hbox{$>$}\ }}
\newcommand{\be}{\begin{equation}}
\newcommand{\ee}{\end{equation}}
\newcommand{\ba}{\begin{eqnarray}}
\newcommand{\ea}{\end{eqnarray}}

\newcommand {\kpcc} {$h_{70}^{-1}$ kpc}

\newcommand {\h} {$h_{70}^{-1}$ Mpc$\;$}
\newcommand {\hh} {$h_{70}^{-1}$ Mpc}
\newcommand {\hhh} {\;h_{70}^{-1} \mathrm{Mpc}}
\newcommand {\ks} {km~s$^{-1} \;$}
\newcommand {\kss} {km~s$^{-1}$}

\newcommand {\mquii} {$\times 10^{15}\;h_{70}^{-1}\;M_{\odot}$}

\newcommand{\degree}{\ensuremath{\mathrm{^\circ}}}
\newcommand{\arcm}{\ensuremath{\mathrm{^\prime}\;}}
\newcommand{\arcs}{\ensuremath{\arcmm\hskip -0.1em\arcmm \;}}
\newcommand{\arcmm}{\ensuremath{\mathrm{^\prime}}}
\newcommand{\arcss}{\ensuremath{\arcmm\hskip -0.1em\arcmm}}

\begin{document}
   \title{Structure of Abell 1995 from optical and X-ray data: a galaxy cluster with an elongated radio halo}  

   \author{
W. Boschin\inst{1,2}
          \and
M. Girardi\inst{2,3}
          \and
R. Barrena\inst{4,5}
}

   \offprints{W. Boschin, \email{boschin@tng.iac.es}}

   \institute{ 
     Fundaci\'on Galileo
     Galilei - INAF (Telescopio Nazionale Galileo),
     Rambla Jos\'e Ana Fern\'andez Perez 7, E-38712 Bre\~na Baja\\
     (La Palma), Canary Islands, Spain\\ 
\and
     Dipartimento di Fisica dell'Universit\`a degli Studi
     di Trieste - Sezione di Astronomia, via Tiepolo 11, I-34143
     Trieste, Italy\\ 
\and INAF - Osservatorio Astronomico di Trieste,
     via Tiepolo 11, I-34143 Trieste, Italy\\ 
\and Instituto de
     Astrof\'{\i}sica de Canarias, C/V\'{\i}a L\'actea s/n, E-38205 La
     Laguna (Tenerife), Canary Islands, Spain\\ 
\and Departamento de
     Astrof\'{\i}sica, Universidad de La Laguna, Av. del
     Astrof\'{\i}sico Francisco S\'anchez s/n, E-38205 La Laguna
     (Tenerife), Canary Islands, Spain\\
}

\date{Received  / Accepted }

\abstract{Abell 1995 is a puzzling galaxy cluster hosting a powerful
  radio halo, but it has not yet been recognized as a obvious cluster
  merger, as usually expected for clusters with diffuse radio
  emission.}{We aim at an exhaustive analysis of the internal
  structure of Abell 1995 to verify if this cluster is really
  dynamically relaxed, as reported in previous studies.}{We base our
  analysis on new and archival spectroscopic and photometric data for
  126 galaxies in the field of Abell 1995. The study of the hot
  intracluster medium was performed on X-ray archival data.}{Based on
  87 fiducial cluster members, we have computed the average cluster
  redshift $\left<z\right>=0.322$ and the global radial velocity
  dispersion $\sigma_{\rm V}\sim 1300$ \kss. We detect two main
  optical subclusters separated by 1.5\arcm that cause the known NE-SW
  elongation of the galaxy distribution and a significant velocity
  gradient in the same direction. As for the X-ray analysis, we
  confirm that the intracluster medium is mildly elongated, but we
  also detect three X-ray peaks. Two X-ray peaks are offset with
  respect to the two galaxy peaks and lie between them, thus
  suggesting a bimodal merger caught in a phase of post core-core
  passage.  The third X-ray peak lies between the NE galaxy peak and a
  third, minor galaxy peak suggesting a more complex merger.  The
  difficulty of separating the two main systems leads to a large
  uncertainty on the line--of-sight (LOS) velocity separation and the
  system mass: $\Delta V_{\rm rf,LOS}=600$--2000 \ks and $M_{\rm
    sys}=2$--5 \mquii, respectively. Simple analytical arguments
  suggest a merging scenario for Abell 1995, where two main subsystems
  are seen just after the collision with an intermediate projection
  angle.}{The high mass of Abell 1995 and the evidence of merging
  suggest it is not atypical among clusters with known radio
  halos. Interestingly, our findings reinforce the previous evidence
  for the peculiar dichotomy between the dark matter and galaxy
  distributions observed in this cluster.}

\titlerunning{Structure of the cluster Abell 1995} \maketitle
%
%________________________________________________________________

\section{Introduction}
\label{intr}

In the past decades, multiwavelength observations from ground and from
space have dramatically shown the complexity of the physical phenomena
occurring in galaxy clusters. An intriguing aspect of these
observations is the discovery of a growing number of clusters
exhibiting diffuse radio emission (on Mpc scale), i.e. large-scale
areas of radio emission without any obvious galaxy counterpart
(Giovannini \& Feretti \cite{gio02}; Ferrari et al. \cite{fer08};
Venturi \cite{ven11}). Particularly prominent are the radio features
known as radio halos, which usually pervade the central cluster
regions in a similar way to the intracluster medium (hereafter
ICM). Instead, radio emission areas found at the edges of clusters are
known as radio relics.

The cause of radio halos and relics is still under investigation. They
are likely to result from synchrotron nonthermal radiation originating
from relativistic electrons of the ICM moving in large-scale cluster
magnetic fields. From a theoretical point of view, cluster mergers
have been proposed as the key process for shedding light on the origin
of these diffuse radio sources. In fact, the huge energy of these
events could reaccelerate mildly relativistic electrons to
relativistic energies and amplify the cluster magnetic fields (e.g.,
Feretti \cite{fer99}). In particular, radio relics seem to be
connected with large-scale shock waves occurring during mergers (e.g.,
Ensslin et al. \cite{ens98}; Hoeft et al. \cite{hoe04}). Instead,
radio halos are probably related to the turbulent motions of the ICM
following a merger (e.g., Cassano et al. \cite{cas06}; Brunetti et
al. \cite{bru09}), but the precise scenario is still being debated.

X-ray observations have been crucial to deriving the dynamical state
of clusters hosting diffuse radio emission. Several statistical
studies (see, e.g., Buote \cite{buo02}; Cassano et al.  \cite{cas10};
Rossetti et al. \cite{ros11}) have discovered interesting correlations
between the properties of radio halos and relics and the ICM X-ray
luminosity and temperature (Giovannini \& Feretti \cite{gio02} and
refs. therein). This is also true when comparing point-to-point the
X-ray and radio surface brightnesses (Govoni et
al. \cite{gov01}). Nevertheless, in a pilot study using the
Sunyaev-Zel'dovich effect, Basu (\cite{bas12}) finds the lack of
bimodality in the radio power -- integrated SZ effect measurement
diagram, the contrary of what is found in the radio power -- X-ray
luminosity diagram (Brunetti et al. \cite{bru07}). This study shows
the need to adopt more investigation techniques in addition to the
X-ray data analysis.

Optical observations can be very helpful when checking for mergers in
clusters with diffuse radio emission and studying their internal
dynamics, too (e.g., Girardi \& Biviano \cite{gir02}). In particular,
combined X-ray/optical studies can be very effective at revealing and
quantifying the level of substructure, when checking for premerging
subsystems and/or merger remnants. The power of this approach comes
from the fact that mergers affect the ICM and the galaxy distributions
in different ways, as shown by numerical simulations (e.g., Roettiger
et al. \cite{roe97}). Thus, X-ray and optical observations complement
each other to provide a more complete picture of merger events.

It is with this scientific rationale in mind that we have begun a
long-term optical observational program to investigate the properties
of clusters hosting radio halos and/or relics: the DARC (``dynamical
analysis of radio clusters'') program (see Girardi et
al.~\cite{gir10conf}). Among the dozens of clusters with known diffuse
radio sources, we decided to perform an optical and X-ray
investigation of the interesting cluster \object{Abell 1995}
(hereafter A1995).

In the X-ray band, A1995 appears as a luminous and hot cluster:
$L_\mathrm{X}$(0.1--2.4 keV)=13.42$\times 10^{44} \ h_{50}^{-2}$
erg\ s$^{-1}$ (B\"ohringer et al. \cite{boe00}); $kT_{\rm X}=7-9$ keV
(from Chandra data, see e.g. Baldi et al. \cite{bal07}, Bonamente et
al. \cite{bon08}, and Ehlert \& Ulmert \cite{ehl09}).

At optical wavelengths, A1995 is a rich cluster (Abell richness class
$=1$; Abell et al. \cite{abe89}). Its light distribution is quite
elongated in the NE-SW direction, but the mass shows a more circular
distribution and is more concentrated than galaxies, as shown by the
weak gravitational lensing reconstruction by Dahle et
al. (\cite{dah02}).  Holhjem et al. (\cite{hol09}) confirm this
interesting discrepancy between light and mass distribution (see
discussion in Sect.~\ref{disc}).

As for previous redshift data, Patel et al. (\cite{pat00}) obtained
spectra for 15 member galaxies and estimate a cluster redshift of
$z=0.322\pm0.001$ and a radial velocity dispersion $\sigma_{\rm
  V}=1282_{-120}^{+153}$ \kss. Irgens et al. (\cite{irg02}) confirm
these measurements on the basis of 20 (unpublished) redshifts, six of
them in common with those of Patel et al. (\cite{pat00}), with
$z=0.3207\pm0.0001$ and $\sigma_{\rm V}=1130_{-110}^{+150}$
\kss. Moreover, they find good agreement between the galaxy velocity
dispersion and the dark matter velocity dispersion obtained from the
weak gravitational lensing analysis ($\sigma_{\rm DM}=1240\pm80$
\kss).

The X-ray ROSAT-HRI emission is peaked on a central bright galaxy and
shows modest elongation in the NE-SW direction, which is not clearly
separated from the emission of two very bright pointlike sources (see
Fig.~1 of Patel et al. \cite{pat00}, see also Ota \& Mitsuda
\cite{ota04}). Ota \& Mitsuda (\cite{ota04}) classify A1995 as a
regular cluster from the stability of the X-ray centroid position and
the good fit with a single $\beta$-model profile. However, despite its
regular appearance, A1995 has a very large cooling time ($t_{\rm
  cool}=10.7$ Gyr; see Fig.~12 of Ota \& Mitsuda \cite{ota04}) and
thus no evidence of a cool core.

Summarizing previous results, there is some hint but no clear evidence
of substructure in A1995, and indeed, on the basis of optical and
X-ray appearance, Pedersen \& Dahle (\cite{ped07}) include this cluster
in the sample of relaxed systems.

\begin{figure*}
\centering 
\includegraphics[width=18cm]{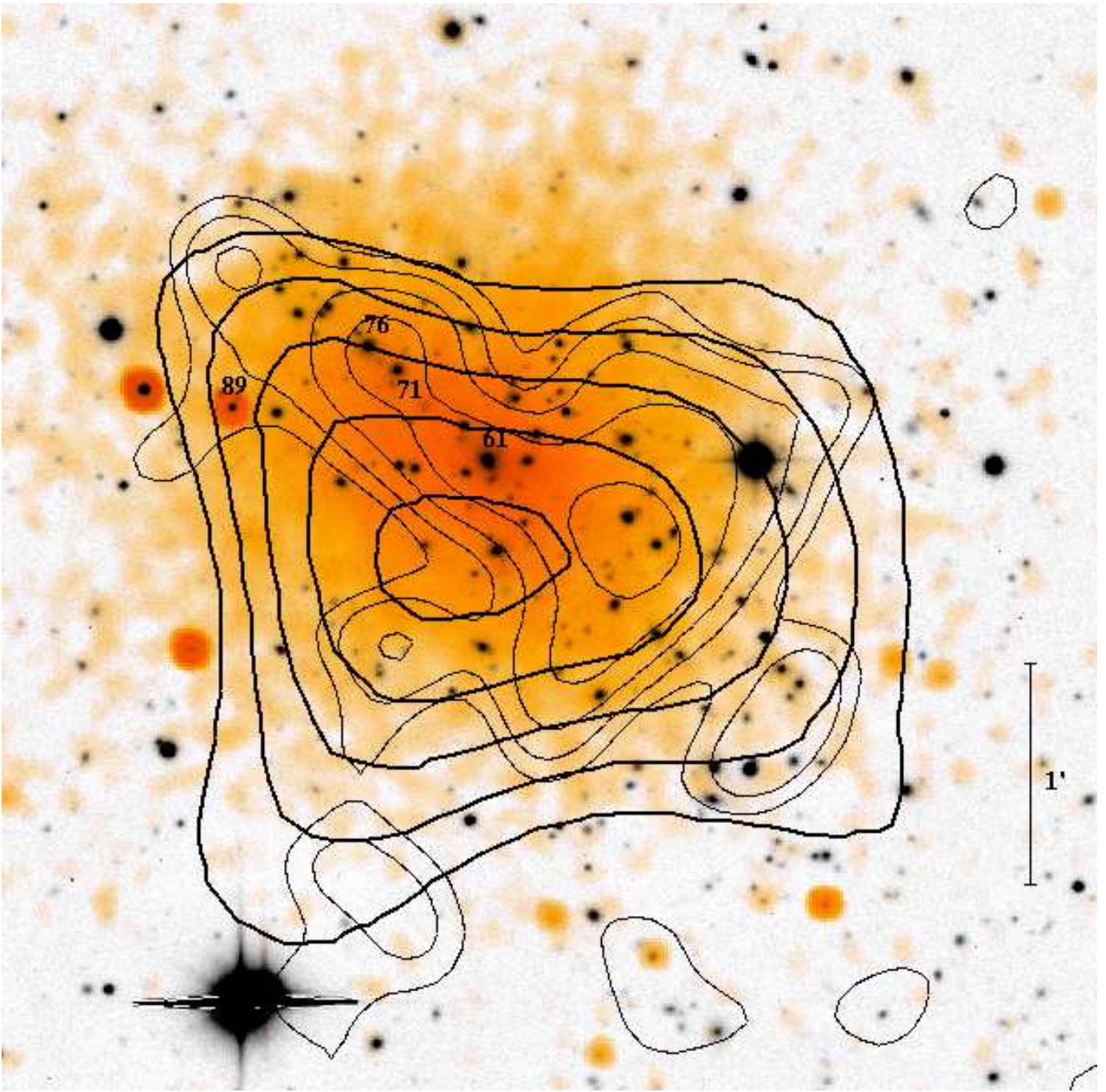}
\caption{Optical/X-ray/radio view of A1995 (direct sky view with north
  up). A smoothed Chandra 0.3-7 keV image (orange and yellow colors)
  of A1995 is superimposed on an $R_{\rm H}$-band image taken with the
  INT. Thin contours are the radio contour levels from Giovannini et
  al. (\cite{gio09}; VLA data at 1.4 GHz with discrete sources
  subtracted, courtesy of F. Govoni). Thick contours are the mass
  distribution contours from Holhjem et al. (\cite{hol09}). Numbers
  highlight notable galaxies mentioned in the text.}
\label{figimage}
\end{figure*}

About the radio wavelengths, Owen et al. (\cite{owe99}) first reported
a possible detection of a diffuse radio source in A1995.  Giovannini
et al. (\cite{gio09}) analyzed new VLA data and discovered an evident
radio halo in this cluster, with a size of $\sim 0.8$ \h and radio
power of $P_{\rm 1.4\,GHz}=1.3\times10^{24}\ h_{70}^{-2}$
W\ Hz$^{-1}$. The radio halo appears somewhat elongated in the NE-SW
direction (Giovannini et al. \cite{gio09}; see also
Fig.~\ref{figimage}).

In the context of our DARC program, we proposed new spectroscopic
observations of A1995 with the Telescopio Nazionale Galileo (TNG). We
also performed new photometric observations at the Isaac Newton
Telescope (INT) and used archival data of the Sloan Digital Sky Survey
(SDSS). As for the analysis in the X-ray band, we used archival
data downloaded from the Chandra Archive.

In this paper, Sect.~2 presents the new spectroscopic and photometric
data. Section~3 describes the analysis of the optical data, while
Sect.~4 presents the analysis of the Chandra archival data. Finally,
in Sect.~5, we discuss our results and propose a scenario for the
dynamical status of A1995.

Throughout this paper, we adopt a flat cosmology with $H_0=70$ km
s$^{-1}$ Mpc$^{-1}$ ($h_{70}=H_0/70$ km s$^{-1}$ Mpc$^{-1}$),
$\Omega_0=0.3$ and $\Omega_{\Lambda}=0.7$. With this choice of the
cosmological parameters, the scale is $\sim 280$ \kpcc/arcmin at the
redshift of A1995. Errors are given at the 68\% confidence level
(hereafter c.l.), unless otherwise stated.

\begin{figure*}
\centering 
\includegraphics[width=18cm]{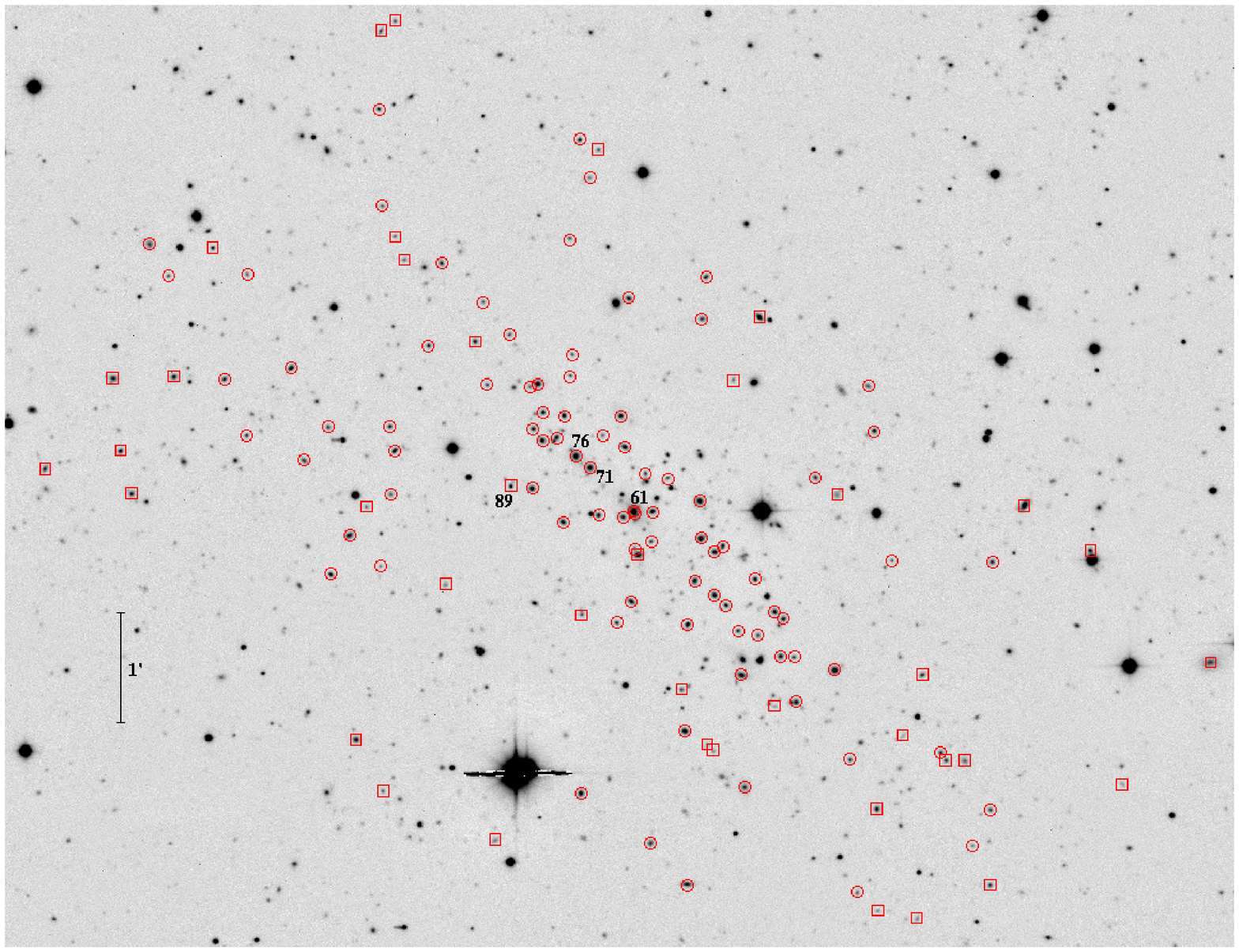}
\caption{Field of the cluster A1995 (direct sky view with north up)
  observed with the INT in the $R_{\rm H}$-band. Cluster members are
  indicated by circles, while nonmember galaxies are shown by squares
  (see Table~\ref{catalogA1995}). Numbers highlight the IDs of notable
  galaxies as in Fig.~\ref{figimage}.}
\label{figottico}
\end{figure*}

\section{Galaxy data and catalog}
\label{data}

\subsection{Spectroscopic observations}
\label{spec}

We performed spectroscopic observations of A1995 in May 2009. As usual
for the clusters in our DARC sample, we used the instrument
DOLORES@TNG\footnote{see http://www.tng.iac.es/instruments/lrs} in MOS
mode with the LR-B grism, which covers the wavelength range 3000--8430
\AA. In summary, we obtained 143 spectra from four observed masks. For
each mask, the total exposure time was 5400 s.

Reduction of spectra and radial velocities computation with the
cross-correlation technique (CC; Tonry \& Davis \cite{ton79}) were
performed using standard IRAF\footnote{see http://iraf.net} tasks, as
for other clusters included in our DARC sample (for a detailed
description see, e.g., Boschin et al. \cite{bos12}, hereafter B12). In
eight cases (IDs.~02, 03, 08, 14, 16, 29, 52, and 91; see
Table~\ref{catalogA1995}) the redshift was estimated with the EMSAO
package (based on the wavelength location of emission lines in the
spectra). Our spectroscopic catalog lists a total of 126 galaxies.

After comparing the velocity measurements for galaxies observed with
multiple masks (see discussion in, e.g., Boschin et al. \cite{bos04},
Girardi et al. \cite{gir11}), we corrected the velocity errors
provided by the CC technique by multiplying them for a factor
$\sim$2. After taking the above correction into account, the median
value of the $cz$ errors is 74 \kss.

We also compared our data with those of Patel et al. (\cite{pat00}),
finding 13 out of 15 galaxies in common. The two redshift measurements
agree with a one--to--one relation, but the $\chi^2$ of the fit
reaches a reasonable value only by multiplying their errors by at
least a factor 1.5. This leads to their typical errors being three
times larger than our typical errors. Considering these large errors,
we preferred not to add this additional information and to study our
(homogeneous) sample.

\subsection{Photometric observations}
\label{int}

We performed photometric observations of A1995 in January 2008 by
using the Wide Field Camera (WFC\footnote{see
  http://www.ing.iac.es/Astronomy/instruments/wfc/}) mounted on the
2.5m INT Telescope. The sky conditions were photometric and the seeing
was $\sim$1.4\arcss. In particular, we observed with the Harris
$B_{\rm H}$ (15 exposures of 600 s) and $R_{\rm H}$ (18 exposures of
300 s) filters. This means a total exposure time of 9000 s and 5400 s
in each band.

We reduced the data and produced our photometric galaxy catalog by
using standard procedures (see, e.g., B12 for details on the reduction
of the WFC images). After transformation of $B_{\rm H}$ and $R_{\rm
  H}$ magnitudes into the $B$ and $R$ Johnson-Cousins magnitudes
(Johnson \& Morgan \cite{joh53}; Cousins \cite{cou76}) and magnitude
correction for the galactic extinction (with $A_B \sim 0.06$ and $A_R
\sim 0.04$, respectively; Schlegel et al. \cite{sch98}), we estimated
that the completeness of the photometric sample is 90\% for
$R\leq$ 21.2 and $B\leq$ 22.7.

\subsection{Galaxy catalog and notable galaxies}
\label{cat}

Table~\ref{catalogA1995} collects all the spectroscopic and
photometric information for the 126 galaxies with a measured redshift
(see also Fig.~\ref{figottico}): ID and IDm (Cols.~1 and 2) are the
identification number of each galaxy and member galaxies,
respectively; Col.~3 reports the equatorial coordinates, $\alpha$ and
$\delta$ (J2000); Cols.~4 and 5 list the $B$ and $R$ magnitudes;
Col.~6 lists the radial (heliocentrically corrected) velocites,
$v=cz_{\sun}$, with their errors, $\Delta v$ (Col.~7).

\input{19508tab1.tex}

A1995 hosts a dominant galaxy (ID.~61, $R=17.78$, hereafter BCG) $\sim
0.8$ mag brighter than the second brightest member galaxy
(ID.~76). The X-ray ROSAT-HRI emission peaks on the BCG (e.g., Patel
et al. \cite{pat00}).

There are several X-ray and radio emitting galaxies in the field of
A1995. Among them, our ID.~89 (a background galaxy) is an evident
pointlike source in Chandra archival data (see
Fig.~\ref{figimage}). Cooray et al. (\cite{coo98}) highlight two
bright radio pointlike sources in A1995: 1453+5803 and
1452+5801. Taking a look at the radio map provided by Giovannini et
al. (\cite{gio09}, see their Fig.~8, right panel), we identify
1453+5803 with our ID.~71, which is a cluster member. Instead,
1452+5801 is likely a background source.

\section{Analysis of the optical data}
\label{anal}

\subsection{Member selection}
\label{memb}

\begin{figure}
\centering
\resizebox{\hsize}{!}{\includegraphics{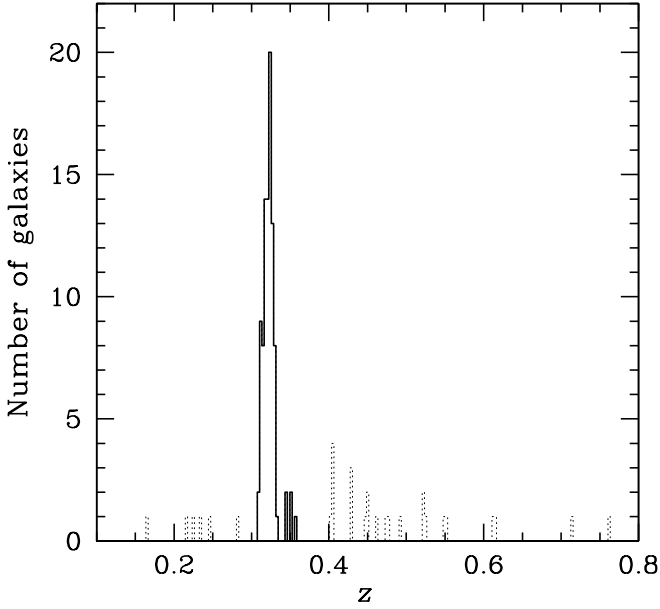}}
\caption
{Histogram of 126 galaxy redshifts. The 94 (provisional) cluster
  members are highlighted with the solid line (see text).}
\label{fighisto}
\end{figure}

As for other DARC clusters, we selected cluster members by running two
statistical tests. First, we used the 1D-DEDICA method (Pisani
\cite{pis93} and \cite{pis96}) on the 126 galaxies with redshifts. The
method detects A1995 as a significant peak (at $>$99\% c.l.) in the
velocity distribution located at $z\sim0.322$. The peak includes 94
(provisional) member galaxies (see Fig.~\ref{fighisto}).

\begin{figure}
\centering
\resizebox{\hsize}{!}{\includegraphics{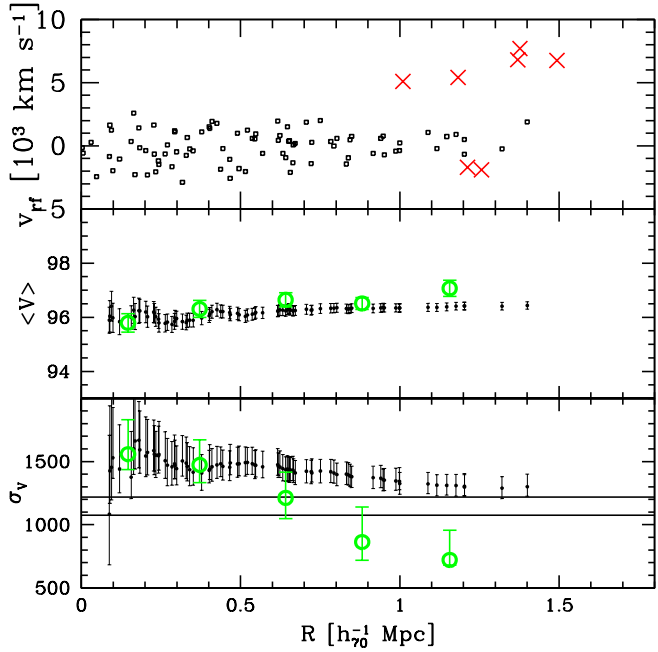}}
\caption
{The 94 (provisional) cluster members (see also Fig.~\ref{fighisto})
  in the phase space (see {\em top panel}). The ordinate is the
  rest-frame velocity, the abscissa the projected clustercentric
  distance. Galaxies rejected by the ``shifting gapper'' procedure are
  shown with (red) crosses. For the cluster center we adopt the
  location of the BCG (see text). Big green circles and small points
  show the differential and integral profiles of the mean velocity (in
  the {\em middle panel}) and the radial velocity dispersion (in the
  {\em bottom panel}). In the bottom panel, two horizontal lines mark
  the range of possible values for the ICM temperature (7-9 keV) with
  their respective errors transformed to $\sigma_{\rm V}$ (see
  Sect.~\ref{disc} for details).}
\label{figprof}
\end{figure}

Then, we used a second tool for member selection, which uses both the
spatial and velocity information: the ``shifting gapper'' method
(Fadda et al. \cite{fad96}; see also, e.g., Girardi et
al. \cite{gir11} for details on the application of this
technique). Here, we only point out that the method needs the
definition of a cluster center. For A1995, we chose the location of
the BCG (see Table~\ref{catalogA1995}). The application of the
``shifting gapper'' rejected another seven galaxies in the outskirts
of the cluster (Fig.~\ref{figprof} -- top panel). Finally we obtained
a sample of 87 fiducial cluster members (Fig.~\ref{figstrip}).

\subsection{Global cluster properties}
\label{glob}

The first and second moments of a distribution can be efficiently
computed by using the biweight estimators of location and scale (Beers
et al. \cite{bee90}). Their application to the velocity distribution
of our 87 cluster members provided a measurement of the mean cluster
redshift and of the global radial velocity dispersion, where we found
$\left<z\right>=0.3217\pm$ 0.0005 (i.e.
$\left<v\right>=96\,437\pm$140 \kss) and $\sigma_{\rm
  V}=1302_{-71}^{+107}$ \kss, respectively.

Analysis of the velocity dispersion profile (Fig.~\ref{figprof})
suggests that the value computed for $\sigma_{\rm V}$ is quite
robust. In fact, the integral profile is asymptotically flat within
the errors. Instead, the decrease in the differential profile agrees
with the fact that the cluster is free of interlopers, as also seen in
the velocity vs. clustercentric distance diagram (Fig.~\ref{figprof} -
top panel). As for the mean velocity profile, the modestly higher
values of the mean velocity in the external region (Fig.~\ref{figprof}
- middle panel) probably come from the slightly larger sampling in the
NE external regions, where galaxies have higher velocities (see
Sect.~\ref{3d}).

\subsection{Analysis of the velocity distribution}
\label{velo}

Deviations of the velocity distribution from Gaussianity can provide
signs of disturbed dynamics. This can be checked by applying classical
shape estimators (Bird \& Beers \cite{bir93}). We did not not find
significant evidence of departures from Gaussianity by using the
skewness, the kurtosis, and the STI estimators.

\begin{figure}
\centering 
\resizebox{\hsize}{!}{\includegraphics{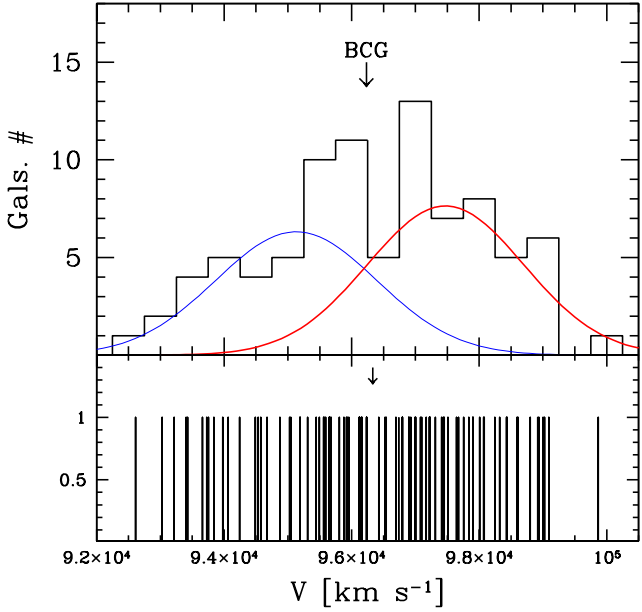}}
\caption
{The 87 galaxies recognized as cluster members.  {\em Top panel}:
  radial velocity distribution with the arrow indicating the velocity
  of the BCG. Red and blue Gaussians are the best two-group partition
  fits according to the 1D-KMM test (see text). {\em Bottom panel}:
  Stripe density plot. The position of the significant gap is marked
  by an arrow.}
\label{figstrip}
\end{figure}

We also searched for significant gaps in the velocity distribution. In
particular, we performed the weighted gap analysis (Beers et
al. \cite{bee91} and \cite{bee92}). We detected one significant gap
that divides the cluster into two groups (see Fig.~\ref{figstrip}):
GV1 (with 42 galaxies and lower velocities) and GV2 (with 45 galaxies
and higher velocities). The BCG resides in the GV1 group, but it lies
on the border with GV2.

\begin{figure}
\centering
\resizebox{\hsize}{!}{\includegraphics{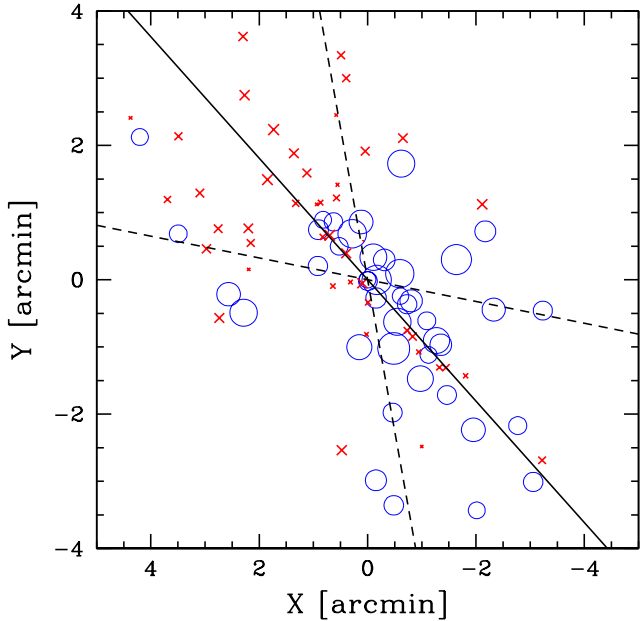}}
\caption
{2D distribution of 87 cluster members. Galaxies with smaller (larger)
  symbols have higher (lower) radial velocities. Blue circles and red
  crosses identify galaxies of GV1 and GV2.  The origin of the
  coordinates is the location of the BCG.  The solid and dashed lines
  indicate the position angle of the cluster velocity gradient (see
  text) and relative errors, respectively.}
\label{figgrad}
\end{figure}

When considering the spatial distributions of the galaxies of GV1 and
GV2, we found that they are different at the 99.94\% c.l. according to
the 2D KS-test (Fasano \cite{fas87}, see our Fig.~\ref{figgrad}). A
statistical test useful to search for eventual subsets in the velocity
distribution is the 1D-Kaye's mixture model test (1D-KMM; Ashman et
al. \cite{ash94}; see also, e.g., B12 for a description of the
method).

We applied this technique by assessing whether a two-Gaussian partition
(accordingly to the detection of the two groups GV1 and GV2) can
provide a significantly better fit to the velocity distribution than a
sole Gaussian. The result of the 1D-KMM test is negative. However,
the best-fit result of the 1D-KMM method gives two groups (KMM1D-1 and
KMM1D-2) of 40 and 47 galaxies, very similar to the groups GV1 and GV2
(but note that the BCG is now belonging to KMM1D-2, the high-velocity
group).  The spatial distributions of the galaxies of KMM1D-1 and
KMM1D-2 are different, too (at the 99.75\% c.l.).  In
Fig.~\ref{figstrip} and Table~\ref{tabsub} we present the results for
the two Gaussians with the velocity dispersions computed by the 1D-KMM
procedure, which considers the membership probabilities of the
galaxies to belong to a group, rather than the velocity dispersions
computed on the galaxy group populations after the assignment. In this
way we could bypass the artificial truncation of the velocity
distribution tails of KMM1D-1 and KMM1D-2, thus minimizing the danger
of heavily underestimating the velocity dispersions of the two
subclusters.

\input{19508tab2.tex}

\subsection{Analysis of the galaxy spatial distribution}
\label{2d}

\begin{figure}
%\centering
\includegraphics[width=8cm]{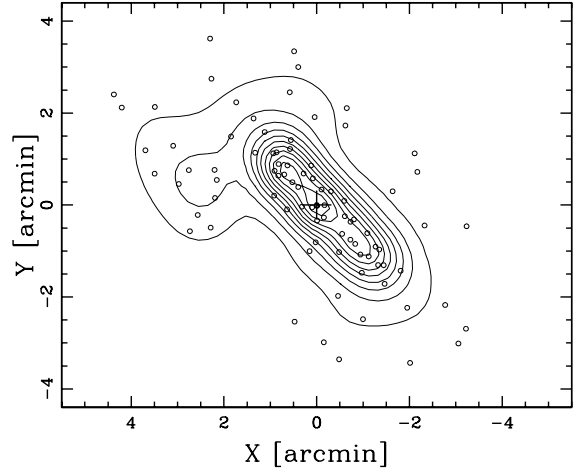}
\caption
{Isodensity contours of the cluster members spatial distribution
  computed with 2D-DEDICA. The origin of the coordinates is the
  location of the BCG (the big cross). Small circles mark the
  positions of cluster members relative to the BCG.}
\label{figk2z}
\end{figure}

We applied the 2D-DEDICA method to the sky positions of A1995 member
galaxies to search for eventual subsets in the galaxy spatial
distribution. We found a NE-SW elongated structure with two
significant peaks, one of them (the NE one) centered on the BCG
(Fig.~\ref{figk2z}). However, this finding could be affected by the
magnitude incompleteness of our spectroscopic sample. To test the
robustness of the 2D-DEDICA results, we resorted to the photometric
catalogs.

\begin{figure}
\centering
\resizebox{\hsize}{!}{\includegraphics{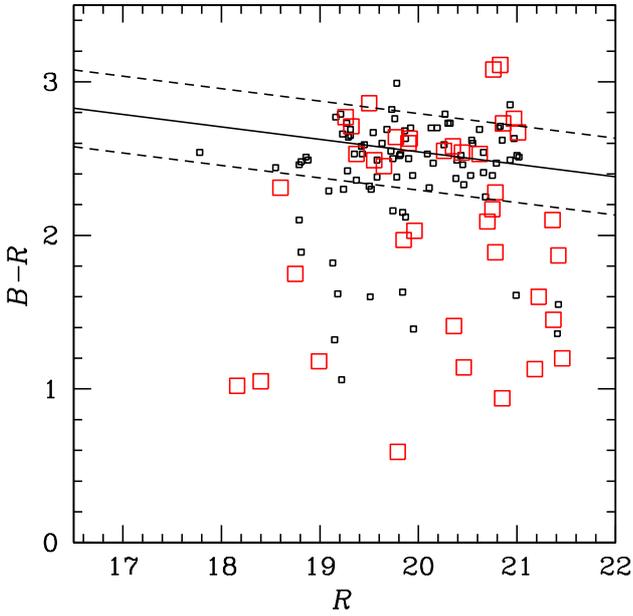}}
\caption
{$B$--$R$ vs. $R$ diagram for the 126 galaxies with spectroscopic
  information. Small black squares are cluster members, while big red
  squares represent field galaxies. The solid line shows the best--fit
  CMR as computed from cluster members; the dashed lines mark the
  region where likely cluster members were selected from the
  photometric catalogs.}
\label{figcmRBR}
\end{figure}

\begin{figure}
\centering
\resizebox{\hsize}{!}{\includegraphics{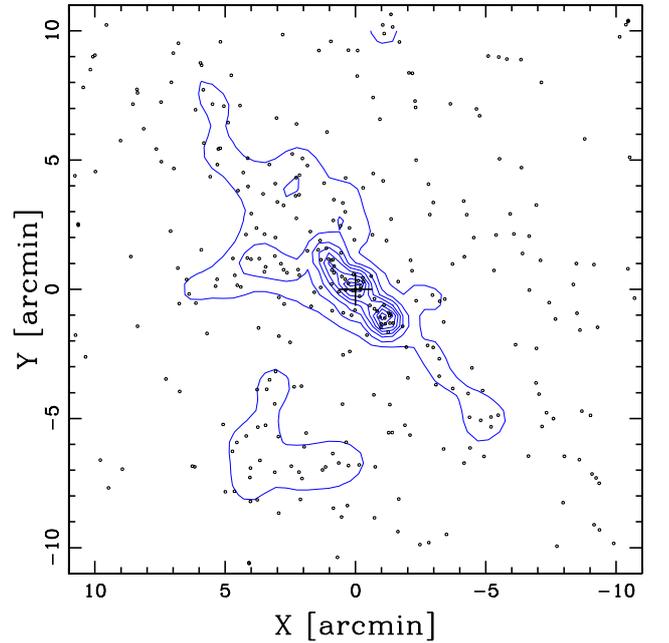}}
\caption
{Results of the 2D-DEDICA method (blue isodensity contour lines)
  applied to likely A1995 members (see text) with $R<$21. The origin
  of the coordinates is the location of the BCG (the big cross). Small
  circles mark the positions of likely cluster members relative to the
  BCG.}
\label{figk2RB}
\end{figure}

We used the color-magnitude relation (hereafter CMR) of early-type
galaxies (the dominant galaxy population in the densest, internal
cluster regions; e.g. Dressler \cite{dre80}) to select likely cluster
members from our photometric sample (see B12 for details on the
technique used for the determination of the CMR and the selection of
member galaxies). We found $B$-$R=4.164-0.081\times R$ (see
Fig.~\ref{figcmRBR}). Figure~\ref{figk2RB} illustrates the contour map
for the likely cluster members (513 with $R<$21 in the whole INT
field): we confirm that A1995 is described well by an elongated
structure. The two most significant peaks have similar densities, one
centered on the BCG galaxy and one $\sim$ 1.5\arcmin at SW.

Table~\ref{tabdedica2d} lists information for these two main peaks
(2D-NE and 2D-SW clumps), including $N_{\rm S}$, the number of members
(Col.~2), the peak densities, $\rho_{\rm S}$, relative to the densest
peak (Col.~4), and the significance of the peaks estimated through the
value of $\chi^2$ (Col.~5). We also detect a minor peak (2D-NENE in
Table~\ref{tabdedica2d}, the third in galaxy density) whose
statistical significance, although nominally acceptable, is much lower
than the two main peaks.

\input{19508tab3.tex}

We also considered the SDSS photometric catalogs (already corrected
for Galactic absorption).  In this case, we selected likely member
galaxies by considering the CMRs in the ($r^{\prime}$--$i^{\prime}$
vs. $r^{\prime}$) and ($g^{\prime}$--$r^{\prime}$ vs. $r^{\prime}$)
color-$r^{\prime}$ mag diagrams (see Goto et al. \cite{got02} and
B12). The two CMRs are $r^{\prime}$--$i^{\prime}$=1.009-0.022$\times$
$r^{\prime}$ and $g^{\prime}$--$r^{\prime}$=2.878-0.062$\times$
$r^{\prime}$, respectively. The result for $r^{\prime}<21.5$ (the
limit for the SDSS star/galaxy classification) confirms the results
reported above.

\subsection{3D analysis: combining spatial and velocity information}
\label{3d}

The 3D tests searching for correlations between positions and
velocities of member galaxies are powerful tools for revealing real
substructures in clusters. First, we checked for a velocity gradient
in the set of the 87 fiducial cluster members (see, e.g., den Hartog
\& Katgert \cite{den96} and Girardi et al. \cite{gir96}). We found a
significant (at the $96\%$ c.l.) velocity gradient with
$PA=48_{-36}^{+33}$ degrees (counter--clock--wise from north). This
means that the NE region of the cluster is populated by
higher velocity galaxies (see Fig.~\ref{figgrad}).

A classical test to detect the presence of substructures is the
$\Delta$-statistics (Dressler \& Schectman \cite{dre88}; hereafter
DS-test). We applied this test by defining, for each galaxy, the
$\delta$ parameter (determined considering the $N_{\rm nn}$=10
neighbors of each galaxy; see, e.g., B12 for a description of the
method) and computing the departure $\Delta$ of the local kinematic
quantities from the global parameters. We also applied the
$\epsilon$-test (Bird~\cite{bir93}) and the $\alpha$-test (West \&
Bothun \cite{wes90}; see also, e.g., Pinkney et al. \cite{pin96} and
Ferrari et al. \cite{fer03} for details of these 3D tests). Both the
DS-test and the $\alpha$-test detected the presence of substructures
(at the $98.5\%$ and $95\%$ c.l., respectively).

A better interpretation of the results of the DS-test can be reached
by splitting the contributions of the local mean velocity (estimator
$\delta_{\rm V}$) and dispersion (estimator $\delta_{\rm s}$) to the
classical $\delta$ parameter (see B12 for details; see also Girardi et
al. \cite{gir97}; Girardi et al. \cite{gir10}). Moreover, we
investigated the results of the DS-test obtained by changing the
number of neighbors.

\begin{figure}
\centering 
\resizebox{\hsize}{!}{\includegraphics{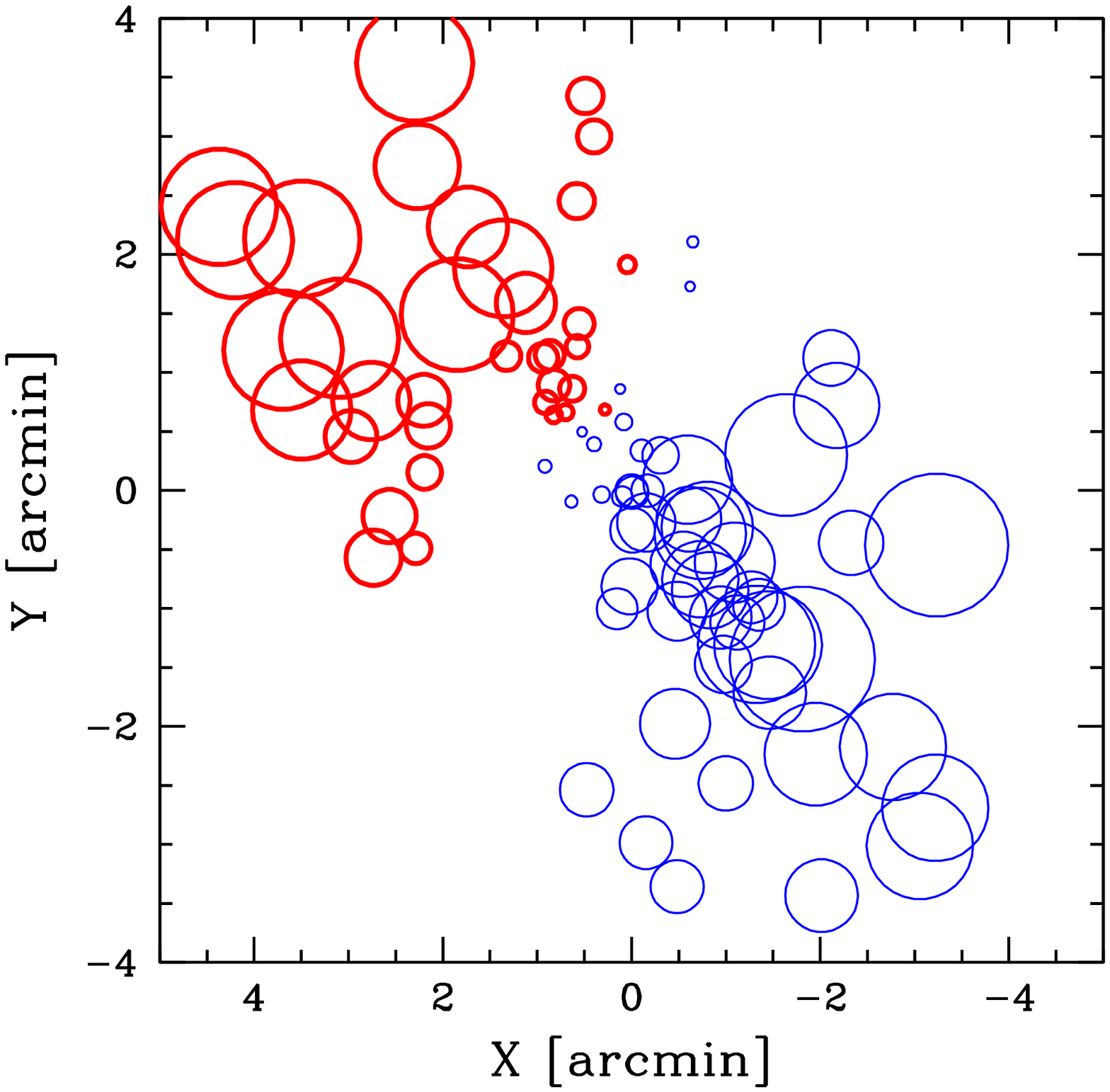}}
\resizebox{\hsize}{!}{\includegraphics{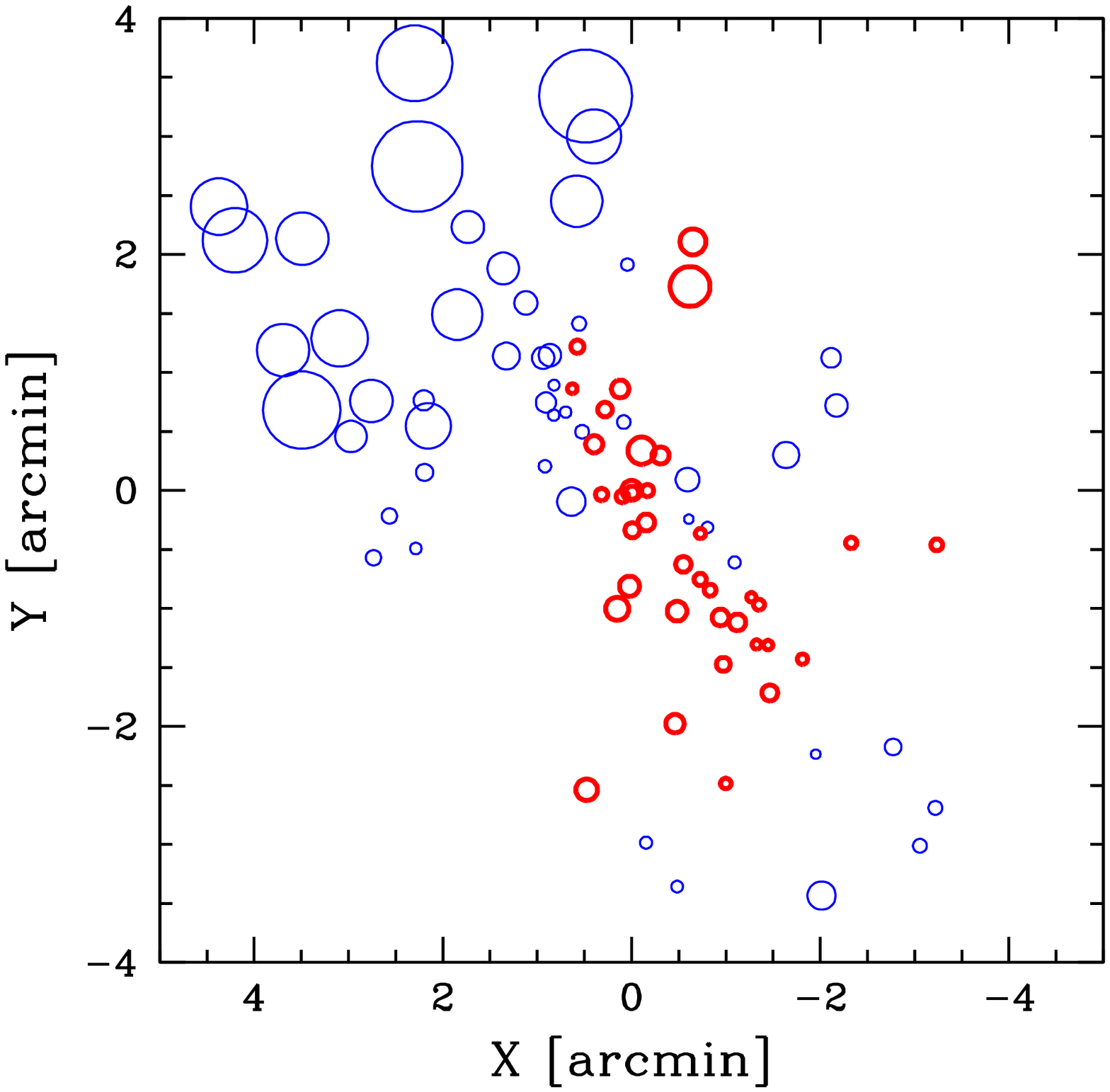}}
\caption
{ Positions on the plane of the sky of 87 cluster members (marked by
  bubbles).  {\em Top panel:} DS-test bubble plot for the deviation
  $\delta_{{\rm V},i}$ and $N_{\rm{nn}}=40$ (see text for an
  explanation). {\em Bottom panel:} as above, but for the deviation
  $\delta_{{\rm s},i}$ and $N_{\rm{nn}}=10$. In both panels, thin/blue
  (thick/red) bubbles show regions with a local value lower (higher)
  than the global one.}
\label{figds}
\end{figure}

The two panels of Fig.~\ref{figds} illustrate the significant results
for the two kinematical indicators. This figure shows the spatial
distribution of the 87 cluster members. Each galaxy is indicated by a
bubble, where the size of the bubble is related to the value of
$\delta$, the deviation of the local kinematic estimator from the
global cluster value.  As for the $\mathrm{\delta_V}$ estimator, the
substructure significance increases up to $>99.9\%$ c.l. when $N_{\rm
  nn}$ =40.  Figure~\ref{figds} (upper panel) shows that there are two
regions of different local mean velocities at NE and SW, in agreement
with the velocity gradient discussed above, and likely corresponding
to a low- and a high-velocity galaxy populations.  When considering
the $\mathrm{\delta_s}$ estimator, the substructure significance is
higher at smaller $N_{\rm nn}$.  Figure~\ref{figds} (lower panel)
shows a NE region of low local velocity dispersions ($N_{\rm nn}$ =10,
significance at the $96\%$ c.l.). Probably, in that region, there is a
minor mixing of the low- and high-velocity populations.

\begin{figure}
\centering
\resizebox{\hsize}{!}{\includegraphics{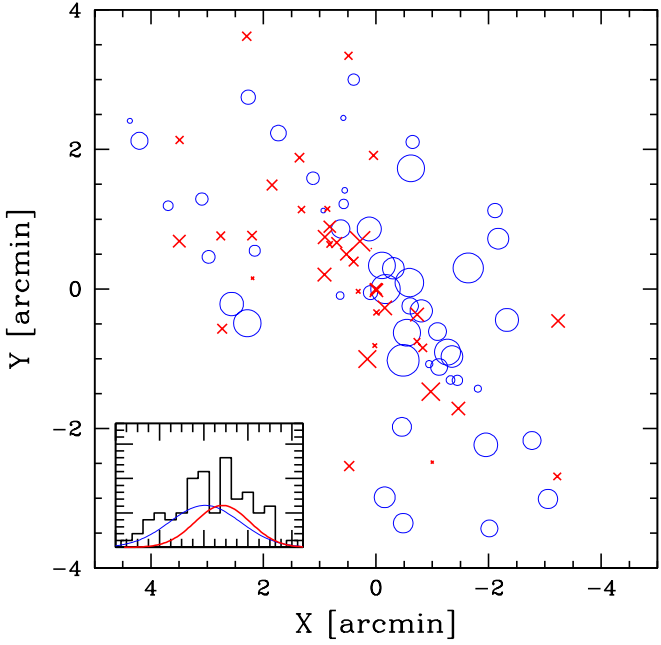}}
\caption
{2D distribution of 87 cluster members. Galaxies with smaller (larger)
  symbols have higher (lower) radial velocities. Big blue circles and
  red crosses indicate galaxies of KMM3D-1 and KMM3D-2.  The insert
  plot shows the same velocity distribution of Fig.~\ref{figstrip} and
  the two Gaussians corresponding to the mean velocity and velocity
  dispersions of KMM3D-1 and KMM3D-2 (blue thin line and red thick
  line, respectively).}
\label{figkmm}
\end{figure}

We also applied the full 3D-KMM method. Considering GV1 and GV2 as a
guess for an initial two-group partition, we find that the 3D galaxy
distribution is well described by a partition of 49 and 38 galaxies
(KMM3D-1 and KMM3D-2 groups). The improvement over the sole 3D
Gaussian is significant at the 98$\%$ c.l. Figure~\ref{figkmm}
illustrates the distribution of KMM3D-1 and KMM3D-2 on the plane of
the sky and the two Gaussians in the velocity
distribution. Table~\ref{tabsub} lists the main properties of these
two groups. With this 3D test, the values of the velocity dispersions
of KMM3D-1 and KMM3D-2 are much higher than those of the corresponding
groups obtained with the 1D methods previously described.

\begin{figure}
\centering 
\resizebox{\hsize}{!}{\includegraphics{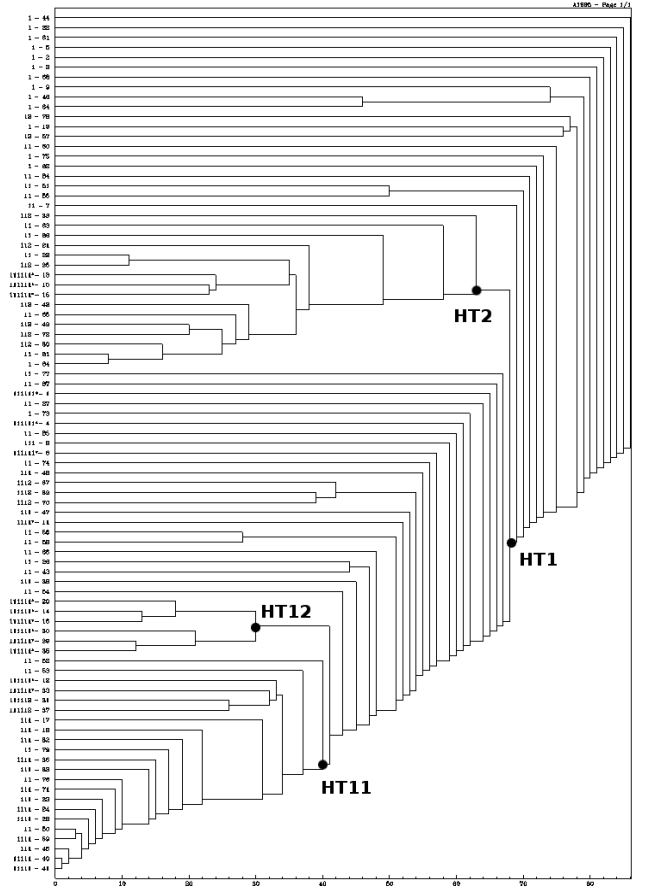}}
%\resizebox{\hsize}{!}{\includegraphics{a1995gerbal.ps}}
\caption
{Results obtained with the Htree method (see text). In this dendogram,
  the horizontal axis reports the binding energy while the vertical
  axis shows the IDms of the member galaxies (as in
  Table~\ref{catalogA1995}).}
\label{a1995gerbal}
\end{figure}

\begin{figure}
\centering
\resizebox{\hsize}{!}{\includegraphics{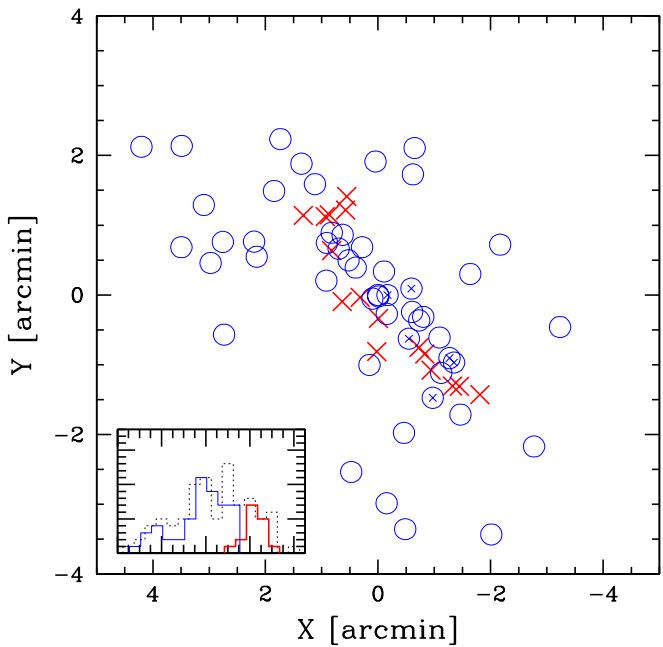}}
\caption
{2D distribution of 87 cluster members.  Big blue circles and red
  crosses indicate galaxies of HT1 (main system) and HT2
  (high-velocity subcluster).  The insert plot shows the same velocity
  distribution of Fig.~\ref{figstrip} (dashed line) and that of HT1
  (blue thin line) and HT2 (red thick line). The subcluster HT12 of
  HT1 is indicated by small blue crosses in the main plot and this
  causes the low-velocity tail in the insert plot.}
\label{fight}
\end{figure}

We finally applied the Htree method (Serna \& Gerbal
\cite{ser96}). This method performs a hierarchical clustering analysis
and returns the relationship between cluster members based on their
relative binding energies (see also, e.g., B12 and Girardi et
al. \cite{gir11}). Figure~\ref{a1995gerbal} illustrates the results of
the Htree method. In this dendogram the binding energy is reported in
abscissa (in arbitrary units) and galaxy pairs and subgroups lie to
the left (at lower energy levels). Going down from the top of the
dendogram, we note the secondary system HT2 and the main system
HT1. HT2 is a group at high velocity. HT1 is the main system and shows
a low-velocity substructure (HT12), while its core (HT11) is dominated
by the BCG. The spatial distributions of galaxies of HT1, HT2, and
HT12 are plotted in Fig.~\ref{fight}.  When using the results of the
Htree method as seeds for the 3D-KMM method, we do not find any two or
three-group partition significantly better than the sole 3D-Gaussian.

\section{X-ray morphological analysis}
\label{Xmorph}

We studied the morphological properties of the ICM by using archival
X-ray data taken with Chandra ACIS--S (exposure ID \#906, total
exposure time 58 ks). We reduced the data using the package
CIAO\footnote{see http://asc.harvard.edu/ciao/} (ver. 4.2) on the chip
S3 in a standard way (see, e.g., Boschin et al. \cite{bos04}).

\begin{figure}
\centering
\resizebox{\hsize}{!}{\includegraphics{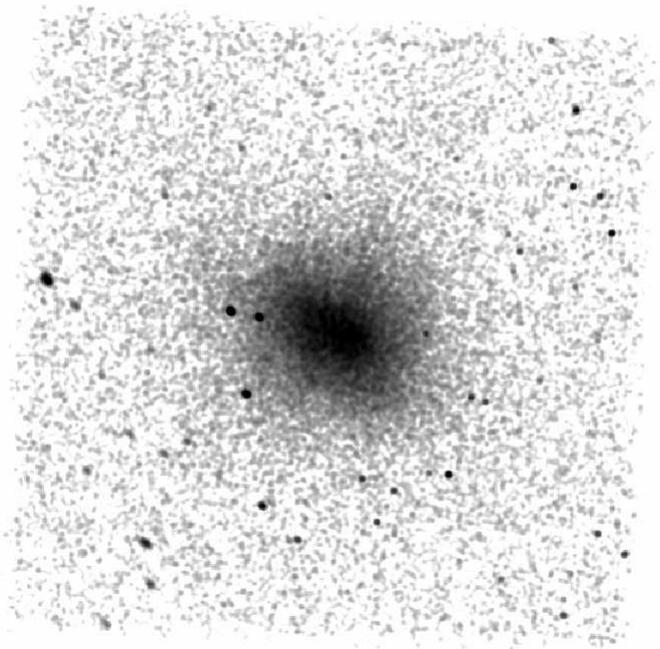}}
\caption
{Chandra image of A1995 (smoothed X-ray emission in the energy band
  0.3--7 keV, direct sky view with north up). The field of view is
  8.5\arcmm$\times$8.5\arcmm.}
\label{figX}
\end{figure}

The reduced image (see Fig.~\ref{figX}) reveals that the ICM exhibits
a regular morphology. From a quantitative point of view, this result
is supported by the power-ratio (Buote \& Tsai \cite{buo96}) analysis
performed by Hart (\cite{har08}). Moreover, we used the task
CIAO/Wavdetect on chip S3 to perform a wavelet multiscale analysis and
found no evidence of multiple clumps in the ICM.

We also used the CIAO package Sherpa to fit an elliptical
$\beta$-model profile (e.g. Boschin et al. \cite{bos04}) to the X-ray
photon distribution (after removing of the pointlike sources found with
CIAO/Wavdetect). The best-fit model is characterized by a centroid
position located $\sim$4$\arcsec$ east of the BCG and a core radius
$r_0=47.3\pm$1.2 arcsec (221$\pm$6 \kpcc). The best-fit values for
other parameters are: $\epsilon=0.213\pm$0.007 (ellipticity),
$\alpha=1.74\pm$0.04 (power law index), and $\theta$=149.8$\pm$1.1
degrees (angle of ellipticity).

\begin{figure}
\centering
\resizebox{\hsize}{!}{\includegraphics{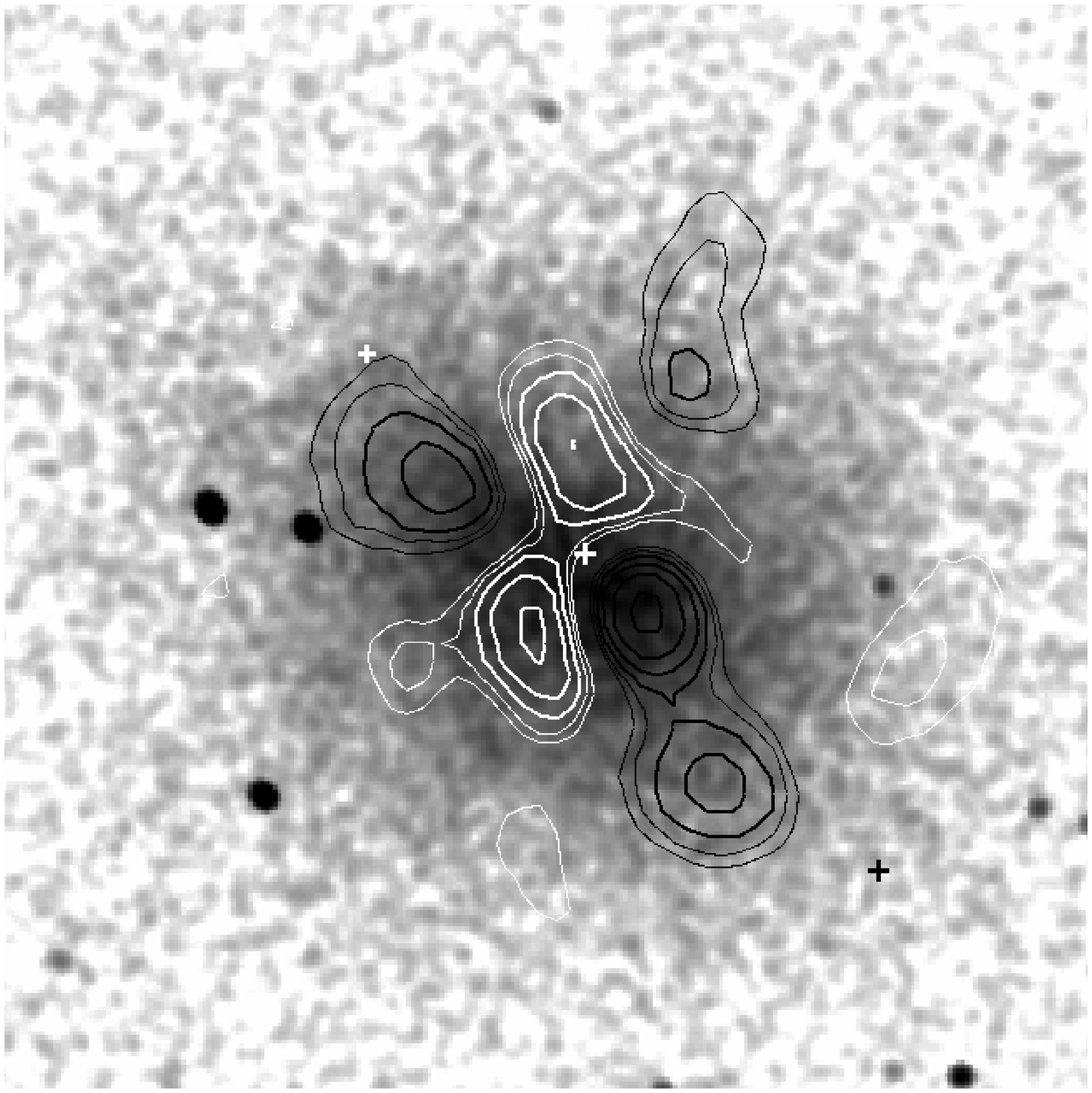}}
\caption
{The smoothed X--ray emission of A1995 (direct sky view with north
  up). Black (white) contours show the positive (negative) elliptical
  $\beta$-mode residuals. Big crosses indicate the two main optical
  clumps detected with the 2D-DEDICA method (2D-NE and 2D-SW in
  Table~\ref{tabdedica2d}). Small cross indicates the third densest
  peak we detect (2D-NENE).}
\label{residui}
\end{figure}

The reduced CSTAT statistic (Cash \cite{cas79}) of the fit is
1.22. Thus, the elliptical beta model is an adequate description to
the data. However, we checked for possible deviations in the X--ray
photon distribution from the above model by computing the model
residuals. We find a deficiency of photons in a dumbbell-shaped region
extending along the SSE--NNW direction in proximity of the X-ray
centroid position (see Fig.~\ref{residui}). Regions with positive
residuals elongated in the NE--SW direction are also found all around
the cluster center.

\section{Discussion and conclusions}
\label{disc}

Our estimate of the velocity dispersion ($\sigma_{\rm
  V}=1302_{-71}^{+107}$ \kss) agrees with those of Patel et al.
(\cite{pat00}) and Irgens et al.  (\cite{irg02}) computed on a much
smaller galaxy sample. This high value of $\sigma_{\rm V}$ predicts
$kT_{\rm X}=10.3_{-1.1}^{+1.4}$ keV (assuming energy equipartition
between galaxies and gas energy per unit mass), and thus agrees with
the high measured X-ray temperature $kT_{\rm X}=7-9$ keV (from Chandra
data, see Baldi et al. \cite{bal07}, Bonamente et al. \cite{bon08},
and Ehlert \& Ulmert \cite{ehl09}). Both $\sigma_{\rm V}$ and $T_{\rm
  X}$ suggest that A1995 is a massive galaxy cluster.

In the hypothesis of dynamical equilibrium (but see in the following)
and typical assumptions (cluster sphericity, galaxies and mass
following identical distributions), we computed virial global
quantities. By considering Girardi \& Mezzetti (\cite{gir01}; see also B12
for details), we obtained a measurement of the mass within the virial
radius $R_{\rm vir}$: $M(<R_{\rm vir}=2.7 \hhh)=3.0_{-0.8}^{+0.9}$
\mquii.

\subsection{Cluster structure and mass}
\label{reali}

Substructure is detected using several analyses of the cluster galaxy
population. Our optical analyses indicate the presence of two main
subclusters aligned in the NE-SW direction causing the velocity
gradient towards the NE direction and separated by $\sim 1.5$\arcmin,
i.e. a projected linear distance $D\sim 0.4$ \hh.

The two subclusters, as determined through 1D- or 3D-KMM methods, have
comparable velocity dispersions within the errors, and in both cases
they are likely two massive systems. The strong uncertainty in the
subcluster membership reflects on the recomputation of the system mass
by summing the masses of the two subclumps; i.e.,  $M_{\rm sys}\sim
2$-5 \mquii, where the low (high) value is computed for the 1D (3D)
case with a rest-frame velocity separation of $\Delta V_{\rm
  rf,LOS}\sim 2000$ \ks ($\sim 600$ \kss; see Table~\ref{tabsub}).
Both 1D and 3D methods have their drawbacks. As for the 1D case, the
two peaks are not clearly separated in the velocity distribution.  In
the 3D case, our spectroscopic sample is not spatially complete and
less extended than the supposed $R_{200}$. We thus suggest that
intermediate values are closer to the real one. This leads to a mass
value not far from the virial mass previously computed considering the
global velocity dispersion.

Moreover, we also have indications of a more complex structure: the
2D-DEDICA analysis detects a third, minor central galaxy peak aligned
along the NE-SW direction. Finally, the Htree method does not detect
a clear bimodal structure.

For the X-ray data analysis, we reaffirm the existence of an isophotes
elongation. More quantitatively, we find isophotes are elongated with
a position angle $PA=\theta$-90\degree=59.8\degree (where $PA$ is
measured in a clockwise direction from the north; see
Sect.~\ref{Xmorph}).  This $PA$ agrees with the direction of the
velocity gradient and is slightly larger than the $PA$ suggested by
the direction defined by the two main optical subclumps. This scenario
is somewhat suggestive of a cluster merger.

Moreover, while previous studies did not find any direct evidence of
substructure in the X-ray emission (Ota \& Mitsuda \cite{ota04};
Pedersen \& Dahle \cite{ped07}), our detailed analysis of Chandra data
suggests the presence of multiple clumps. The elliptical $\beta$-model
residuals resemble the residuals computed by Girardi et
al. (\cite{gir10}) from the fit of the X-ray photon distribution of
the cluster Abell 2294 (see their Fig.~12) with a simple
$\beta$-model. For Abell 2294, Girardi et al. (\cite{gir10}) explain
the residual image by the presence of two very closely projected
subsystems. Our Fig.~\ref{residui} suggests a similar scenario for
A1995, with an excess of X-ray emission (positive residuals) in the
NE-SW direction. The main central X-ray peak is close to the BCG. This
peak and a SW one are offset with respect to the two main optical
peaks and located between them, thus resembling the usual situation of
a bimodal merging just after the core-core passage with the
collisional X-ray components slowed down with respect to the
collisionless galaxy clumps. The NE X-ray peak lies between the main
optical peak and the third (minor) optical peak, thus suggesting that
a third subcluster intervenes in the merger. A fourth X-ray peak
(excess of positive residuals) at the NW is less significant and is
not considered.  Summarizing, our results confirm the active dynamical
state of A1995 from the thermodynamic point of view.

\subsection{Merger kinematics and diffuse radio sources}
\label{2model}

\begin{figure}
\centering
\resizebox{\hsize}{!}{\includegraphics{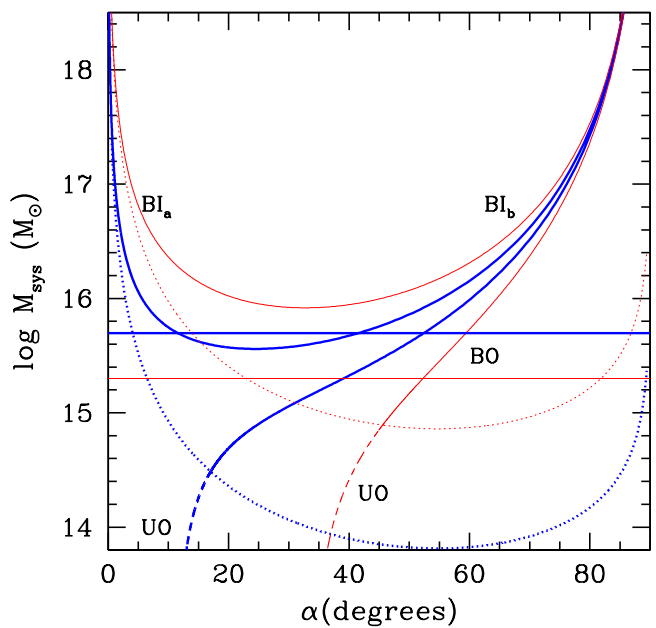}}
\caption
{System mass -- projection angle diagram of the analytic bimodal model
  applied to the two subclusters. Thin/red (thick/blue) lines refer to
  the case-1D (case-3D; see text). In particular, bound (solid curves)
  and incoming (collapsing; BI$_{\rm a}$ and BI$_{\rm b}$), bound
  outgoing (expanding; BO), and unbound (dashed curves) outgoing
  (UO) solutions are shown. The horizontal lines give the computed
  values of the system mass for the two cases.  The dotted curves
  separate bound and unbound regions (above and below the dotted
  curves, respectively).}
\label{figbim}
\end{figure}

Considering only the two main subclusters, the merger can be followed
through the analytic method of the bimodal model (Beers et
al. \cite{bee82}; see also Boschin et al. \cite{bos10}). For the
values of the relevant parameters ($D$; $M_{\rm sys}$; $\Delta V_{\rm
  rf,LOS}$), we used the values reported above taking both the cases
corresponding to 1D- and 3D-KMM analysis into account (hereafter
case-1D and case-3D). We assumed $t=0.2$ Gyr for the time elapsed from
the core crossing (a typical value for clusters hosting radio halos;
e.g. Markevitch et al. \cite{mar02}; Girardi et
al. \cite{gir08}). Figure~\ref{figbim} illustrates the solutions of
the bimodal model, where the mass of the system $M_{\rm sys}$ is
plotted against $\alpha$, the projection angle between the direction
defined by the subclumps and the plane of the sky. In both case-1D and
case-3D we find a bound outgoing solution (BO, see Fig.~\ref{figbim})
compatible with $\alpha \sim 50$\degree. This value agrees with the
strong evidence of substructure in the 3D analyses, since 3D
substructure is more easily detected at intermediate angles (Pinkney
et al. \cite{pin96}).  Note that the assumption of significantly
longer times $t$ would lead to larger, unlikely, angles of view
(e.g. $>70$\degree$\,$ for $t>1$ Gyr).

As a comparison with the results in the radio band, Giovannini et
al. (\cite{gio09}) show that, as usual, the radio halo of A1995 and
the X-ray emitting ICM occupy the cluster volume in the same manner
(see their Fig.~8 and our Fig.~\ref{figimage}).  We find that the
elongation of the radio halo roughly agrees with the direction of the
merger, as detected in several clusters, e.g. the Bullet Cluster
(1E0657-56; Markevitch et al. \cite{mar02}) and Abell 520 (Girardi et
al. \cite{gir08}); but see Abell 523 (Giovannini et al. \cite{gio11}).

In conclusion, A1995, for its high mass and cluster merging evidence,
is not atypical among clusters hosting radio halos. The remaining
puzzling point concerns the dark matter distribution, which is quite
circularly symmetric with respect to the galaxy distribution (Dahle et
al. \cite{dah02}; Holhjem et al. \cite{hol09}), the latter indicating
a NE-SW preferential direction. Owing to the assumed collisionless
nature of both galaxies and dark matter particles, this is surprising
evidence. Moreover, this disagrees with the results found for other
extensively studied clusters, such as the Bullet Cluster (e.g.,
Markevitch et al. \cite{mar02}) or clusters CL 0152-1357 and MS
1054-0321 (Jee et al. \cite{jee05a}, \cite{jee05b}), where a
coincidence is found between the galaxy and dark matter
distributions. Nevertheless, the peculiarity of A1995 is not so
extreme as in other clusters, such as CL 0024+17, where
evidence was found for a dark matter ring (Jee et al. \cite{jee07})
without a obvious galaxy counterpart (Qin et al. \cite{qin08}; but see
also Ponente \& Diego \cite{pon11}, who suggest that dark matter
ringlike structures could be due to systematics in lensing
reconstruction). In any case, the observed dichotomy between the dark
matter and galaxy distributions makes A1995 an appealing target for
future studies.

\begin{acknowledgements}

M.G. acknowledges financial support from ASI-INAF (I/088/06/0 grant)
and PRININAF2010. This work has been supported by the Programa Nacional de
Astronom\'\i a y Astrof\'\i sica of the Spanish Ministry of Science
and Innovation under grants AYA2010-21322-C03-02, AYA2007-67965-C03-01,
and AYA2010-21887-C04-04.  

This publication is based on observations made with the Telescopio
Nazionale Galileo (TNG) and the Isaac Newton Telescope (INT). The TNG
is operated by the Fundaci\'on Galileo Galilei -- INAF. The INT is
operated by the Isaac Newton Group. Both telescopes are located in the
Spanish Observatorio of the Roque de Los Muchachos of the Instituto de
Astrofisica de Canarias (island of La Palma, Spain).

This research made use of the NASA/IPAC Extragalactic Database
(NED), operated by the Jet Propulsion Laboratory (California
Institute of Technology), under contract with the National Aeronautics
and Space Administration.

This research is also partly based on photometric data of the Sloan
Digital Sky Survey (SDSS). Funding for the SDSS has been provided by
the Alfred P. Sloan Foundation, the Participating Institutions, the
National Aeronautics and Space Administration, the National Science
Foundation, the U.S. Department of Energy, the Japanese
Monbukagakusho, and the Max Planck Society. The SDSS Web site is
http://www.sdss.org/.

The SDSS is managed by the Astrophysical Research Consortium for the
Participating Institutions. The Participating Institutions are the
American Museum of Natural History, Astrophysical Institute Potsdam,
University of Basel, University of Cambridge, Case Western Reserve
University, University of Chicago, Drexel University, Fermilab, the
Institute for Advanced Study, the Japan Participation Group, Johns
Hopkins University, the Joint Institute for Nuclear Astrophysics, the
Kavli Institute for Particle Astrophysics and Cosmology, the Korean
Scientist Group, the Chinese Academy of Sciences (LAMOST), Los Alamos
National Laboratory, the Max-Planck-Institute for Astronomy (MPIA),
the Max-Planck-Institute for Astrophysics (MPA), New Mexico State
University, Ohio State University, University of Pittsburgh,
University of Portsmouth, Princeton University, the United States
Naval Observatory, and the University of Washington.
\end{acknowledgements}

\end{document}

%% file: 19508tab1.tex
%\documentclass[referee]{aa}
%\usepackage{graphicx}
%%new commands
%\def\lesssim{\mathrel{\hbox{\rlap{\hbox{\lower4pt\hbox{$\sim$}}}\hbox{$<$}}}}
%\def\gtrsim{\mathrel{\hbox{\rlap{\hbox{\lower4pt\hbox{$\sim$}}}\hbox{$>$}}}}
%\newcommand{\mincir}{\raise -2.truept\hbox{\rlap{\hbox{$\sim$}}\raise5.truept
%\hbox{$<$}\ }}
%\newcommand{\magcir}{\raise -2.truept\hbox{\rlap{\hbox{$\sim$}}\raise5.truept
%\hbox{$>$}\ }}
%\newcommand{\siml}{\raise -2.truept\hbox{\rlap{\hbox{$\sim$}}\raise5.truept
%\hbox{$<$}\ }}
%\newcommand{\simg}{\raise -2.truept\hbox{\rlap{\hbox{$\sim$}}\raise5.truept
%\hbox{$>$}\ }}
%\newcommand{\be}{\begin{equation}}
%\newcommand{\ee}{\end{equation}}
%\newcommand{\ba}{\begin{eqnarray}}
%\newcommand{\ea}{\end{eqnarray}}
%\newcommand {\h} {$h^{-1}$ Mpc $ \;$}
%\newcommand {\kpc} {$h^{-1}$ kpc}
%\newcommand {\hh} {$h^{-1}$ Mpc}
%\newcommand {\ks} {km~s$^{-1} \;$}
%\newcommand {\kss} {km~s$^{-1}$}
%\newcommand {\mpc} {$Mpc \;$}
%\newcommand {\msun} {$h^{-1} \  M_{\odot} \;$}
%\newcommand {\m} {$M_{\odot} \;$}
%\newcommand {\ml} {$h \, M_{\odot}/L_{\odot} \;$}
%\newcommand {\mll} {$h \, M_{\odot}/L_{\odot}$}
%\newcommand{\vel}{\,{\rm km\,s^{-1}}}
%\newcommand{\tng}{\mathrm{T}}
%\newcommand{\sds}{\mathrm{S}}
%\newcommand{\tns}{\mathrm{T+S}}
%%
%\begin{document}

%\addtocounter{table}{-2}
\begin{table}[!ht]
        \caption[]{Catalog of 126 galaxies in the field of A1995 with
          measured radial velocities, where $\dagger$ highlights the
          ID. of the BCG.}
         \label{catalogA1995}
              $$ 
        % \begin{array}{p{0.5\linewidth}l}
           \begin{array}{r c c c c r r}
            \hline
            \noalign{\smallskip}
            \hline
            \noalign{\smallskip}

\mathrm{ID} & \mathrm{IDm} & \mathrm{\alpha},\mathrm{\delta}\,(\mathrm{J}2000)  & B & R& v\,\,\,\,\,&\mathrm{\Delta}v\\
  & & & & &\mathrm{(\,km}&\mathrm{s^{-1}\,)}\\
            \hline
            \noalign{\smallskip}  

 01& -   &14\ 52\ 18.12 ,+58\ 01\ 33.3& 19.18& 18.16&     18112& 101   \\  
 02& -   &14\ 52\ 24.17 ,+58\ 00\ 27.5& 22.31& 21.18&     74182&  82   \\  
 03& -   &14\ 52\ 26.31 ,+58\ 02\ 34.0& 20.50& 18.75&     70357& 170   \\  
 04& -   &14\ 52\ 30.82 ,+58\ 02\ 58.2& 19.45& 18.40&     49107& 129   \\  
 05& 1   &14\ 52\ 33.02 ,+58\ 02\ 27.6& 22.62& 20.09&     95649&  48   \\  
 06& 2   &14\ 52\ 33.15 ,+58\ 00\ 13.9& 22.75& 20.38&     97643&  72   \\  
 07& -   &14\ 52\ 33.19 ,+57\ 59\ 33.2& 22.93& 20.35&    121114& 208   \\  
 08& 3   &14\ 52\ 34.39 ,+57\ 59\ 54.6& 22.77& 21.41&     95562& 100   \\  
 09& -   &14\ 52\ 34.92 ,+58\ 00\ 40.5& 22.54& 19.91&    128719& 141   \\  
 10& -   &14\ 52\ 36.19 ,+58\ 00\ 40.6& 21.77& 20.36&    103177& 105   \\  
 11& 4   &14\ 52\ 36.53 ,+58\ 00\ 44.9& 22.42& 20.11&     95879&  77   \\  
 12& -   &14\ 52\ 37.80 ,+58\ 01\ 26.9& 21.99& 19.96&    138505&  73   \\  
 13& -   &14\ 52\ 38.18 ,+57\ 59\ 15.2& 22.67& 20.78&    134283& 128   \\  
 14& -   &14\ 52\ 39.12 ,+58\ 00\ 54.0& 22.82& 21.22&    143292& 100   \\  
 15& 5   &14\ 52\ 39.87 ,+58\ 02\ 28.6& 23.42& 20.93&     94671&  73   \\  
 16& -   &14\ 52\ 40.85 ,+57\ 59\ 19.3& 23.29& 21.42&    103596& 100   \\  
 17& -   &14\ 52\ 40.89 ,+58\ 00\ 14.5& 22.36& 19.50&    128750&  72   \\  
 18& 6   &14\ 52\ 41.04 ,+58\ 03\ 38.5& 22.07& 19.58&     95191&  41   \\  
 19& 7   &14\ 52\ 41.47 ,+58\ 04\ 02.7& 22.62& 20.15&     96695&  97   \\  
 20& 8   &14\ 52\ 42.26 ,+57\ 59\ 29.3& 23.54& 20.83&     96151&  76   \\  
 21& 9   &14\ 52\ 42.72 ,+58\ 00\ 41.2& 22.95& 20.43&     94533&  85   \\  
 22& -   &14\ 52\ 43.56 ,+58\ 03\ 04.4& 21.82& 19.85&    142517& 117   \\  
 23&10   &14\ 52\ 43.78 ,+58\ 01\ 29.5& 21.29& 18.81&     98437&  65   \\  
 24&11   &14\ 52\ 45.07 ,+58\ 03\ 13.3& 22.27& 19.72&     93027& 120   \\  
 25&12   &14\ 52\ 46.39 ,+58\ 01\ 12.4& 21.82& 19.50&     95659&  88   \\  
 26&13   &14\ 52\ 46.53 ,+58\ 01\ 36.7& 23.31& 20.62&     97678& 116   \\  
 27&14   &14\ 52\ 47.26 ,+58\ 01\ 57.2& 22.04& 19.45&     95027&  49   \\  
 28&15   &14\ 52\ 47.45 ,+58\ 01\ 37.0& 22.52& 19.76&     98072& 108   \\  
 29& -   &14\ 52\ 47.86 ,+58\ 01\ 10.0& 22.66& 21.46&    134784& 100   \\  
 30&16   &14\ 52\ 47.86 ,+58\ 02\ 01.0& 21.73& 19.37&     93980&  45   \\  
 31& -   &14\ 52\ 48.86 ,+58\ 04\ 39.9& 20.17& 18.99&     20008&  68   \\  
 32&17   &14\ 52\ 48.98 ,+58\ 01\ 48.2& 23.16& 20.54&     96159&  85   \\  
 33&18   &14\ 52\ 49.20 ,+58\ 02\ 18.7& 22.50& 19.87&     95919&  85   \\  
 34&19   &14\ 52\ 49.92 ,+58\ 00\ 26.3& 20.80& 19.18&     99092&  87   \\  
 35&20   &14\ 52\ 50.11 ,+58\ 01\ 26.9& 21.37& 18.88&     94063&  57   \\  
 36&21   &14\ 52\ 50.33 ,+58\ 01\ 50.7& 23.05& 20.32&     98429&  77   \\  
 37& -   &14\ 52\ 50.67 ,+58\ 04\ 06.1& 23.84& 20.76&    157123&  60   \\  
 38&22   &14\ 52\ 51.17 ,+58\ 02\ 04.6& 22.37& 19.68&     97314&  81   \\  
 39&23   &14\ 52\ 51.38 ,+58\ 02\ 36.5& 21.99& 19.84&     94876&  57   \\  
 40&24   &14\ 52\ 51.94 ,+58\ 02\ 33.4& 22.00& 19.21&     95569&  49   \\  
 41&25   &14\ 52\ 51.96 ,+58\ 02\ 10.1& 21.93& 19.29&     97901&  48   \\  
 42& -   &14\ 52\ 52.03 ,+58\ 00\ 45.9& 23.46& 21.36&    184420&  68   \\   
  
                        \noalign{\smallskip}			    
            \hline					    
            \noalign{\smallskip}			    
            \hline					    
         \end{array}
     $$ 
         \end{table}
\addtocounter{table}{-1}
\begin{table}[!ht]
          \caption[ ]{Continued.}
     $$ 
           \begin{array}{r r c c c r r}
            \hline
            \noalign{\smallskip}
            \hline
            \noalign{\smallskip}

\mathrm{ID} & \mathrm{IDm} &\mathrm{\alpha},\mathrm{\delta}\,(\mathrm{J}2000)  & B & R& v\,\,\,\,\,&\mathrm{\Delta}v\\ 
  & & & & &\mathrm{(\,km}&\mathrm{s^{-1}\,)}\\ 

            \hline
            \noalign{\smallskip}

 43  & - &14\ 52\ 52.44 ,+58\ 00\ 49.1& 22.82  &21.37    & 214244&  61   \\ 
 44          &26 &14\ 52\ 52.54 ,+58\ 05\ 01.7& 22.33  &19.81    &  97005&  61   \\ 
 45          &27 &14\ 52\ 52.77 ,+58\ 04\ 39.0& 21.34  &19.95    &  93754&  72   \\ 
 46          &28 &14\ 52\ 52.87 ,+58\ 02\ 40.7& 20.89  &18.79    &  96237&  40   \\ 
 47          &29 &14\ 52\ 52.97 ,+58\ 03\ 00.8& 20.70  &18.81    &  93425&  61   \\ 
 48          &30 &14\ 52\ 53.31 ,+58\ 02\ 17.7& 21.96  &19.43    &  93727&  49   \\ 
 49          &31 &14\ 52\ 53.81 ,+57\ 59\ 33.9& 21.54  &19.24    &  95494& 180   \\ 
 50          &32 &14\ 52\ 53.81 ,+58\ 01\ 53.9& 20.28  &19.22    &  92613&  41   \\ 
 51          &33 &14\ 52\ 53.98 ,+58\ 00\ 56.7& 20.47  &19.15    &  95648&  69   \\ 
 52  & - &14\ 52\ 54.19 ,+58\ 01\ 19.0& 21.60  &20.46    &  64859& 100   \\ 
 53          &34 &14\ 52\ 55.13 ,+58\ 03\ 13.1& 23.47  &20.85    &  95045&  56   \\ 
 54          &35 &14\ 52\ 56.18 ,+58\ 02\ 55.2& 21.70  &19.28    &  93215&  49   \\ 
 55          &36 &14\ 52\ 56.28 ,+58\ 02\ 39.0& 23.03  &20.30    &  95310& 105   \\ 
 56          &37 &14\ 52\ 56.30 ,+57\ 59\ 56.2& 21.47  &19.84    &  95189&  89   \\ 
 57          &38 &14\ 52\ 56.66 ,+58\ 03\ 15.5& 23.15  &20.55    &  93840&  72   \\ 
 58  & - &14\ 52\ 57.17 ,+58\ 02\ 31.8& 20.91  &18.60    &  84327&  64   \\ 
 59          &39 &14\ 52\ 57.39 ,+58\ 02\ 35.1& 22.60  &20.99    &  98081& 109   \\ 
 60          &40 &14\ 52\ 57.41 ,+58\ 02\ 54.1& 22.77  &19.78    &  95675&  84   \\ 
 61          &\dagger 41 &14\ 52\ 57.46 ,+58\ 02\ 55.3& 20.32  &17.78    &  96231&  43   \\ 
 62          &42 &14\ 52\ 57.62 ,+58\ 02\ 06.6& 21.89  &19.23    &  98609&  51   \\ 
 63          &43 &14\ 52\ 57.79 ,+58\ 04\ 50.1& 21.96  &19.58    &  97228&  85   \\ 
 64          &44 &14\ 52\ 58.08 ,+58\ 03\ 30.1& 22.00  &19.27    &  99868&  96   \\ 
 65          &45 &14\ 52\ 58.20 ,+58\ 02\ 52.2& 22.21  &19.54    &  96795&  45   \\ 
 66          &46 &14\ 52\ 58.37 ,+58\ 03\ 46.9& 21.38  &19.09    &  94487&  44   \\ 
 67          &47 &14\ 52\ 58.61 ,+58\ 01\ 55.1& 22.91  &20.45    &  94242&  48   \\ 
 68          &48 &14\ 52\ 59.59 ,+58\ 03\ 36.4& 23.53  &21.02    &  93405& 169   \\ 
 69          &49 &14\ 52\ 59.86 ,+58\ 02\ 53.2& 22.55  &19.73    &  98597&  89   \\ 
 70  & - &14\ 52\ 59.88 ,+58\ 06\ 10.3& 23.94  &20.83    & 147730&  88   \\ 
 71          &50 &14\ 53\ 00.46 ,+58\ 03\ 18.8& 21.25  &18.79    &  96896&  72   \\ 
 72          &51 &14\ 53\ 00.46 ,+58\ 05\ 55.3& 23.26  &20.79    &  97437&  84   \\ 
 73  & - &14\ 53\ 01.01 ,+58\ 02\ 00.4& 22.79  &20.70    & 156865&  60   \\ 
 74          &52 &14\ 53\ 01.05 ,+58\ 00\ 23.1& 21.93  &19.16    &  96796&  61   \\ 
 75          &53 &14\ 53\ 01.15 ,+58\ 06\ 15.9& 22.62  &19.92    &  97450&  57   \\ 
 76          &54 &14\ 53\ 01.42 ,+58\ 03\ 25.1& 20.99  &18.55    &  95934&  40   \\ 
 77          &55 &14\ 53\ 01.66 ,+58\ 04\ 20.1& 23.52  &21.00    &  98800&  61   \\ 
 78          &56 &14\ 53\ 01.80 ,+58\ 04\ 08.4& 23.78  &20.93    &  97905&  80   \\ 
 79          &57 &14\ 53\ 01.85 ,+58\ 05\ 22.4& 22.93  &20.68    &  98920&  69   \\ 
 80          &58 &14\ 53\ 02.21 ,+58\ 03\ 47.0& 21.82  &19.52    &  95808&  53   \\ 
 81          &59 &14\ 53\ 02.28 ,+58\ 02\ 49.6& 22.00  &19.42    &  98320&  64   \\ 
 82          &60 &14\ 53\ 02.74 ,+58\ 03\ 35.1& 21.90  &19.74    &  96526& 113   \\ 
 83          &61 &14\ 53\ 03.67 ,+58\ 03\ 48.8& 22.40  &19.90    &  96121&  85   \\ 
 84          &62 &14\ 53\ 03.70 ,+58\ 03\ 33.6& 21.96  &19.31    &  98009&  56   \\  
   
                        \noalign{\smallskip}			    
            \hline					    
            \noalign{\smallskip}			    
            \hline					    
         \end{array}
     $$ 
         \end{table}
\addtocounter{table}{-1}
\begin{table}[!ht]
          \caption[ ]{Continued.}
     $$ 
           \begin{array}{r c c c c r r}
            \hline
            \noalign{\smallskip}
            \hline
            \noalign{\smallskip}

\mathrm{ID} & \mathrm{IDm} & \mathrm{\alpha},\mathrm{\delta}\,(\mathrm{J}2000)  & B & R & v\,\,\,\,\,&\mathrm{\Delta}v\\ 
  & & & & &\mathrm{(\,km}&\mathrm{s^{-1}\,)}\\ 

            \hline
            \noalign{\smallskip}
   
 85  &63 &14\ 53\ 04.01 +58\ 04\ 04.1& 21.37&  18.86&       98245&  56    \\    
 86  &64 &14\ 53\ 04.32 +58\ 03\ 40.0& 22.35&  19.82&       95444&  80    \\    
 87  &65 &14\ 53\ 04.39 +58\ 03\ 07.6& 22.00&  19.31&       95593&  45    \\    
 88  &66 &14\ 53\ 04.54 +58\ 04\ 02.7& 22.79&  20.46&       99003&  84    \\    
 89  & - &14\ 53\ 05.83 +58\ 03\ 09.2& 20.38&  19.79&      228343&  53    \\    
 90  &67 &14\ 53\ 05.93 +58\ 04\ 30.6& 22.92&  20.53&       97168&  68    \\    
 91  & - &14\ 53\ 06.91 +57\ 59\ 57.4& 21.79&  20.85&       67552& 100    \\    
 92  &68 &14\ 53\ 07.49 +58\ 04\ 03.6& 23.20&  20.66&       97757&  69    \\    
 93  &69 &14\ 53\ 07.75 +58\ 04\ 48.2& 23.60&  20.97&       96928& 121    \\    
 94  & - &14\ 53\ 08.28 +58\ 04\ 26.9& 22.50&  19.90&      120968&  45    \\    
 95  & - &14\ 53\ 10.29 +58\ 02\ 15.7& 23.15&  20.62&      183678& 117    \\    
 96  &70 &14\ 53\ 10.58 +58\ 05\ 09.3& 22.54&  19.86&       96426&  60    \\    
 97  &71 &14\ 53\ 11.43 +58\ 04\ 24.7& 22.33&  19.94&       96544&  93    \\    
 98  & - &14\ 53\ 13.13 +58\ 05\ 11.1& 23.59&  20.86&      165411&  56    \\    
 99  &72 &14\ 53\ 13.75 +58\ 03\ 28.1& 22.23&  19.63&       97509&  73    \\    
100  & - &14\ 53\ 13.75 +58\ 05\ 23.7& 23.68&  21.01&      134645&  57    \\    
101  & - &14\ 53\ 13.75 +58\ 07\ 20.1& 23.06&  20.78&      106610& 152    \\    
102  &73 &14\ 53\ 14.04 +58\ 03\ 04.5& 23.06&  20.27&       99024& 164    \\    
103  &74 &14\ 53\ 14.11 +58\ 03\ 41.1& 22.83&  20.13&       96994&  81    \\    
104  & - &14\ 53\ 14.54 +58\ 00\ 24.4& 23.73&  20.97&      121583& 112    \\    
105  &75 &14\ 53\ 14.64 +58\ 05\ 40.1& 23.07&  20.66&       96738&  69    \\    
106  & - &14\ 53\ 14.71 +58\ 07\ 14.7& 22.81&  20.26&      105445& 116    \\    
107  &76 &14\ 53\ 14.74 +58\ 02\ 25.8& 22.97&  21.42&       93661& 144    \\    
108  &77 &14\ 53\ 14.86 +58\ 06\ 32.5& 22.89&  20.19&       97096&  77    \\    
109  & - &14\ 53\ 15.70 +58\ 02\ 57.7& 22.92&  20.75&      164881& 125    \\    
110  & - &14\ 53\ 16.39 +58\ 00\ 52.1& 21.90&  19.37&      121226&  53    \\    
111  &78 &14\ 53\ 16.85 +58\ 02\ 42.3& 20.95&  19.13&       94574&  73    \\    
112  &79 &14\ 53\ 18.12 +58\ 02\ 21.1& 22.24&  19.74&       96899&  55    \\    
113  &80 &14\ 53\ 18.31 +58\ 03\ 40.8& 22.88&  20.39&       97218& 104    \\    
114  &81 &14\ 53\ 19.94 +58\ 03\ 22.7& 21.99&  19.87&       97081& 128    \\    
115  &82 &14\ 53\ 20.86 +58\ 04\ 12.6& 21.88&  19.35&       97227&  64    \\    
116  &83 &14\ 53\ 23.86 +58\ 05\ 03.3& 23.14&  20.75&       97412& 132    \\    
117  &84 &14\ 53\ 23.88 +58\ 03\ 36.2& 22.85&  20.26&       95955& 116    \\    
118  &85 &14\ 53\ 25.39 +58\ 04\ 06.7& 22.16&  19.78&       97836&  73    \\    
119  & - &14\ 53\ 26.21 +58\ 05\ 17.4& 22.98&  20.44&       93916&  65    \\    
120  & - &14\ 53\ 28.85 +58\ 04\ 07.9& 22.04&  19.55&       94197&  68    \\    
121  &86 &14\ 53\ 29.26 +58\ 05\ 02.5& 23.51&  20.81&       96121&  77    \\    
122  &87 &14\ 53\ 30.55 +58\ 05\ 19.6& 21.11&  19.51&       98937& 100    \\    
123  & - &14\ 53\ 31.73 +58\ 03\ 04.7& 22.03&  19.32&      120824& 104    \\    
124  & - &14\ 53\ 32.47 +58\ 03\ 27.8& 22.41&  19.77&      128644&  65    \\    
125  & - &14\ 53\ 33.05 +58\ 04\ 06.8& 22.03&  19.26&      156585&  61    \\    
126  & - &14\ 53\ 37.61 +58\ 03\ 18.2& 22.10&  19.65&      105365&  76    \\    

                        \noalign{\smallskip}			    
            \hline					    
            \noalign{\smallskip}			    
            \hline					    
         \end{array}
     $$ 
\end{table}

%%
%\end{document}

%% file: 19508tab2.tex
\begin{table*}
        \caption[]{Kinematical properties of several subsystems.}
         \label{tabsub}
            $$
         \begin{array}{l r r r r r}
            \hline
            \noalign{\smallskip}
\mathrm{System} & N_{\rm g} &<v>\;\;\;\;&\sigma_{\rm V}\;\;\;& M\;\;\;\;\;\;& \mathrm{Notes}\;\;\;\;\;\;\;\;\;\;\;\;\;\;\;\;\;\\
& &\mathrm{km\ s^{-1}}\;\;&\mathrm{km\ s^{-1}}& 10^{15}h_{70}^{-1}M_{\odot}&\\
         \hline
         \noalign{\smallskip}
\mathrm{Whole\ system} &87& 96437\pm140& 1302_{\ -71}^{+107}& 3.0\;\;\;&\\
\mathrm{KMM1D-1}       &40& 95130^{\mathrm{a}}      &  954^{\mathrm{a}}&1.2\;\;\;&\mathrm{1D\;low\;velocity\;subcluster}\\
\mathrm{KMM1D-2}       &47& 97472^{\mathrm{a}}     &  805^{\mathrm{a}}&0.7\;\;\;&\mathrm{1D\;high\;velocity\;subcluster}\\
\mathrm{KMM3D-1}       &49& 96055\pm200& 1389_{\ -91}^{+122}&3.6\;\;\;&\mathrm{3D\;low\;velocity\;subcluster}\\
\mathrm{KMM3D-2}       &38& 96850\pm178& 1080_{-113}^{+166}&1.7\;\;\;&\mathrm{3D\;high\;velocity\;subcluster}\\
\mathrm{HT1}           &51& 95829\pm134&  950_{-107}^{\ +99}&1.2\;\;\;&\mathrm{main\;system}\\
\mathrm{HT2}           &16& 98200\pm 90&  342_{\ -44}^{\ +76}&0.05\;\;\;&\mathrm{secondary\;(high\;velocity)\;system}\\
              \noalign{\smallskip}
             \noalign{\smallskip}
            \hline
            \noalign{\smallskip}
            \hline
         \end{array}
$$
\begin{list}{}{}  
\item[$^{\mathrm{a}}$] Here we consider $<v>$ and $\sigma_{\rm V}$ as
  given by the KMM software, where galaxies are opportunely weighted
  according to their membership probability (see Sect.~\ref{velo}).
\end{list}
         \end{table*}

%% file: 19508tab3.tex
\begin{table}
        \caption[]{Substructure from analysis of the INT photometric sample.}
         \label{tabdedica2d}
            $$
         \begin{array}{l r c c c }
            \hline
            \noalign{\smallskip}
            \hline
            \noalign{\smallskip}
\,\,\,\,\,\mathrm{Subclump} & N_{\rm S} & \,\,\alpha,\,\delta\,({\rm J}2000)&\rho_{
\rm S}&\chi^2_{\rm S}\\
& & \mathrm{h:m:s,\degree:\arcmm:\arcs}&&\\
         \hline
         \noalign{\smallskip}
\mathrm{2D-SW\ (INT\ }R<21)       & 21&14\ 52\ 48.9+58\ 01\ 48&1.00&20\\
\mathrm{2D-NE\ (INT\ }R<21)       & 23&14\ 52\ 57.6+58\ 03\ 03&0.99&18\\
\mathrm{2D-NENE\ (INT\ }R<21)     & 13&14\ 53\ 04.1+58\ 03\ 50&0.70&11\\
              \noalign{\smallskip}
              \noalign{\smallskip}
            \hline
            \noalign{\smallskip}
            \hline
         \end{array}
$$
         \end{table}

%% file: 19508.bbl
\begin{thebibliography}{}

\bibitem[1989]{abe89} Abell, G. O., Corwin, H. G. Jr., \& Olowin, R. P. 1989, \apjs, 70, 1

\bibitem[1994]{ash94} Ashman, K. M., Bird, C. M., \& Zepf, S. E. 1994, \aj, 108, 2348

\bibitem[2007]{bal07} Baldi, A., Ettori, S.; Mazzotta, P., Tozzi, P., \& Borgani, S. 2007, \apj, 666, 835

\bibitem[2012]{bas12} Basu, K. 2012, \mnras, 421, L112

 
\bibitem[1990]{bee90} Beers, T. C., Flynn, K., \& Gebhardt, K. 1990, \aj, 100, 32

\bibitem[1991]{bee91} Beers, T. C., Forman, W., Huchra, J. P., Jones, C., \& Gebhardt, K. 1991, \aj, 102, 1581

\bibitem[1992]{bee92} Beers, T. C., Gebhardt, K., Huchra, J. P., et al. 1992, \apj, 400, 410


\bibitem[1982]{bee82} Beers, T. C., Geller, M. J., \& Huchra, J. P. 1982, \apj, 257, 23

\bibitem[1993]{bir93} Bird, C. M., \& Beers, T. C. 1993, \aj, 105, 1596


\bibitem[2000]{boe00} B\"ohringer, H., Voges, W., Huchra, J. P., et al. 2000, \apjs, 129, 435

\bibitem[2008]{bon08} Bonamente, M., Joy, M., LaRoque, S. J., et al. 2008, \apj, 675, 106

\bibitem[2010]{bos10} Boschin, W., Barrena, R., \& Girardi, M. 2010, \aap, 521, A78

\bibitem[2004]{bos04} Boschin, W., Girardi, M., Barrena, R., et al. 2004, \aap, 416, 839

\bibitem[2012]{bos12} Boschin, W., Girardi, M., Barrena, R., \& Nonino, M. 2012, \aap, 540, A43 (B12)

\bibitem[2009]{bru09} Brunetti, G., Cassano, R., Dolag, K., \& Setti, G. 2009, \aap, 507, 661

\bibitem[2007]{bru07} Brunetti, G., Venturi, T., Dallacasa, D., et al. 2007, \apj, 670, L5

\bibitem[2002]{buo02} Buote, D. A. 2002, in ``Merging Processes in Galaxy Clusters'', eds. L. Feretti, I. M. Gioia, \& G. Giovannini (The Netherlands, Kluwer Ac. Pub.): Optical Analysis of Cluster Mergers

\bibitem[1996]{buo96} Buote, D. A., \& Tsai, J. C. 1996, \apj, 458, 27

\bibitem[1979]{cas79} Cash, W. 1979, \apj, 228, 939

\bibitem[2006]{cas06} Cassano, R., Brunetti, G., \& Setti, G. 2006, \mnras, 369, 1577

\bibitem[2010]{cas10} Cassano, R., Ettori, S., Giacintucci, S., et al. 2010, \apjl, 721, 82

\bibitem[1998]{coo98} Cooray, A. R., Grego, L., Holzapfel, W. L., et al. 1998, \aj, 115, 1388

\bibitem[1976]{cou76} Cousins, A. W. J., 1976, Mem. R. Astr. Soc, 81, 25

\bibitem[2002]{dah02} Dahle, H., Kaiser, N., Irgens, R. J., Lilje, P. B., \& Maddox, S. J. 2002, \apjs, 139, 313

\bibitem[1996]{den96} den Hartog, R., \& Katgert, P. 1996, \mnras, 279, 349

\bibitem[1980]{dre80} Dressler, A. 1980, \apj, 236, 351

\bibitem[1988]{dre88} Dressler, A., \& Shectman, S. A. 1988, \aj, 95, 985

\bibitem[2009]{ehl09} Ehlert, S., \& Ulmer, M. P. 2009, \aap, 503, 35

\bibitem[1998]{ens98} Ensslin, T. A., Biermann, P. L., Klein, U., \& Kohle, S. 1998, \aap, 332, 395

\bibitem[1996]{fad96} Fadda, D., Girardi, M., Giuricin, G., Mardirossian, F., \& Mezzetti, M. 1996, \apj, 473, 670

\bibitem[1987]{fas87} Fasano, G., \& Franceschini, A. 1987, \mnras, 225, 155

\bibitem[1999]{fer99} Feretti, L. 1999, MPE Report No. 271

\bibitem[2008]{fer08} Ferrari, C., Govoni, F., Schindler, S., Bykov, A. M., \& Rephaeli, Y. 2008, \ssr, 134, 93

\bibitem[2003]{fer03} Ferrari, C., Maurogordato, S., Cappi, A., \& Benoist, C. 2003, \aap, 399, 813

\bibitem[2009]{gio09} Giovannini, G., Bonafede, A., Feretti, L., et al.  2009, \aap, 507, 1257

\bibitem[2002]{gio02} Giovannini, G., \& Feretti, L. 2002, in ``Merging Processes in Galaxy Clusters'', eds. L. Feretti, I. M. Gioia, \& G. Giovannini (The Netherlands, Kluwer Ac. Pub.): Diffuse Radio Sources and Cluster Mergers

\bibitem[2011]{gio11} Giovannini, G., Feretti, L., Girardi, M., et al. 2011, \aap, 530, L5

\bibitem[2011]{gir11} Girardi, M., Bardelli, S., Barrena, R., et al. 2011, \aap, 536, A89

\bibitem[2010]{gir10conf} Girardi, M., Barrena, R., \& Boschin, W. 2010, Contribution to the conference ``Galaxy clusters: observations, physics and cosmology'', held in Garching (Germany), July 26-30 2010. Published online at the site http://www.mpa-garching.mpg.de/$\sim$clust10/

\bibitem[2008]{gir08} Girardi, M., Barrena, R., Boschin, W., \& Ellingson, E. 2008, \aap, 491, 379

\bibitem[2002]{gir02} Girardi, M., \& Biviano, A. 2002, in ``Merging Processes in Galaxy Clusters'', eds. L. Feretti, I. M. Gioia, \& G. Giovannini (The Netherlands, Kluwer Ac. Pub.): Optical Analysis of Cluster Mergers

\bibitem[2010]{gir10} Girardi, M., Boschin, W., \& Barrena, R. 2010, \aap, 517, A65

\bibitem[1997]{gir97} Girardi, M., Escalera, E., Fadda, D., et al. 1997, \apj, 482, 11

\bibitem[1996]{gir96} Girardi, M., Fadda, D., Giuricin, G. et al. 1996, \apj, 457, 61

\bibitem[2001]{gir01} Girardi, M., \& Mezzetti, M. 2001, \apj, 548, 79

\bibitem[2002]{got02} Goto, T., Sekiguchi, M., Nichol, R. C., et al. 2002, \aj, 123, 1807

\bibitem[2001]{gov01} Govoni, F., Ensslin, T. A., Feretti, L., \& Giovannini, G. 2001, \aap, 369, 441

\bibitem[2008]{har08} Hart, B. 2008, PhD Thesis, preprint arXiv:0801.4093

\bibitem[2004]{hoe04} Hoeft, M., Br\"uggen, M., \& Yepes, G. 2004, \mnras, 347, 389

\bibitem[2009]{hol09} Holhjem, K., Schirmer, M., \& Dahle, H. 2009, \aap, 504, 1

\bibitem[2002]{irg02} Irgens, R. J., Lilje, P. B., Dahle, H., \& Maddox, S. J.
2002, \apj, 579, 227

\bibitem[2007]{jee07} Jee, M. J., Ford, H. C., Illingworth, G. D., et al. 2007, \apj, 661, 728 

\bibitem[2005a]{jee05a} Jee, M. J., White, R. L., Ben\'itez, N., et al. 2005a, \apj, 618, 46

\bibitem[2005b]{jee05b} Jee, M. J., White, R. L., Ford, H. C., et al. 2005b, \apj, 634, 813

\bibitem[1953]{joh53} Johnson, H. L., \&  Morgan, W. W. 1953, \apj, 117, 313

\bibitem[2002]{mar02} Markevitch, M., Gonzalez, A. H., David, L., et al. 2002, \apj, 567, L27

\bibitem[2004]{ota04} Ota, N., \& Mitsuda, K. 2004, \aap, 428, 757

\bibitem[1999]{owe99} Owen, F., Morrison, G., \& Voges, W. 1999, proceedings of the workshop ``Diffuse Thermal and Relativistic Plasma in Galaxy Clusters'', eds. H. B\"ohringer, L. Feretti, \& P. Schuecker, MPE Report 271, pp. 9--11

\bibitem[2000]{pat00} Patel, S. K., Joy, M., Carlstrom, J. E., et al. 2000, \apj, 541, 37

\bibitem[2007]{ped07} Pedersen, K., \& Dahle, H. 2007, \apj, 667, 26

\bibitem[1996]{pin96} Pinkney, J., Roettiger, K., Burns, J. O., \& Bird, C. M. 1996, \apjs, 104, 1

\bibitem[1993]{pis93} Pisani, A. 1993, \mnras, 265, 706

\bibitem[1996]{pis96} Pisani, A. 1996, \mnras, 278, 697

\bibitem[2011]{pon11} Ponente, P. P., \& Diego, J. M. 2011, \aap, 535, A119

\bibitem[2008]{qin08} Qin, B., Shan, H.-Y., \& Tilquin, A. 2008, \apj, 679, L81

\bibitem[1997]{roe97} Roettiger, K., Loken, C., \& Burns, J. O. 1997, \apjs, 109, 307

\bibitem[2011]{ros11} Rossetti, M., Eckert, D., Cavalleri, B. M., et al. 2011, \aap, 532, A123

\bibitem[1998]{sch98} Schlegel, D. J., Finkbeiner, D. P., \& Davis, M. 1998, \apj, 500, 525

\bibitem[1996]{ser96} Serna, A., \& Gerbal, D. 1996, \aap, 309, 65

\bibitem[1979]{ton79} Tonry, J., \& Davis, M. 1979, \apj, 84, 1511

\bibitem[2011]{ven11} Venturi, T. 2011, Mem. SAIt, 82, 499

\bibitem[1990]{wes90} West, M. J., \& Bothun, G. D. 1990, \apj, 350, 36

\end{thebibliography}
